\documentclass[12pt, draftcls, onecolumn]{IEEEtran}

\usepackage{color}
\usepackage{graphicx}
\usepackage{amsmath, amsfonts, amssymb, psfrag, epsfig, epstopdf, bbm}
\usepackage{cite}
\usepackage{caption}
\usepackage{todonotes}

\newcommand\scalemath[2]{\scalebox{#1}{\mbox{\ensuremath{\displaystyle #2}}}}

\newcommand{\markov}{\mathrel{\multimap}\joinrel\mathrel{-}%
\joinrel\mathrel{\mkern-6mu}\joinrel\mathrel{-}}

\newcommand{\bm}[1]{\mbox{\boldmath{$#1$}}}
\newtheorem{proposition}{{Proposition}}
\newtheorem{definition}{{Definition}}
\newtheorem{theorem}{{Theorem}}
\newtheorem{lemma}{{Lemma}}
\newtheorem{example}{{Example}}
\newtheorem{corollary}{{Corollary}}

\DeclareMathAlphabet{\mathpzc}{OT1}{pzc}{m}{it}

\newcommand{\mli}[1]{\mathit{#1}}

\ifCLASSOPTIONcompsoc
\usepackage[caption=false,font=normalsize,labelfont=sf,textfont=sf]{subfig}
\else
\usepackage[caption=false,font=footnotesize]{subfig}
\fi

\makeatletter

\newcommand{\xleftrightarrow}[2][]{\ext@arrow 3359\leftrightarrowfill@{#1}{#2}}
\newcommand{\xdashleftrightarrow}[2][]{\ext@arrow 3359\leftrightarrowfill@@{#1}{#2}}
\def\rightarrowfill@@{\arrowfill@@\relax\relbar\rightarrow}
\def\leftarrowfill@@{\arrowfill@@\leftarrow\relbar\relax}
\def\leftrightarrowfill@@{\arrowfill@@\leftarrow\relbar\rightarrow}
\def\arrowfill@@#1#2#3#4{%
  $\m@th\thickmuskip0mu\medmuskip\thickmuskip\thinmuskip\thickmuskip
   \relax#4#1
   \xleaders\hbox{$#4#2$}\hfill
   #3$%
}

\makeatother
\newcounter{parentnumber}
\allowdisplaybreaks

\begin{document}

\title{Two-Way Source-Channel Coding}


\author{
  \IEEEauthorblockN{Jian-Jia Weng,~\IEEEmembership{Student Member,~IEEE}, Fady Alajaji,~\IEEEmembership{Senior Member,~IEEE},\\ and Tam\'as Linder,~\IEEEmembership{Fellow,~IEEE}}
\thanks{%
    The authors are with the Department of Mathematics and Statistics, Queen's University, Kingston, ON K7L 3N6, Canada (Emails: jian-jia.weng@queensu.ca, \{fady, linder\}@mast.queensu.ca).}
\thanks{
  This work was supported in part by NSERC of Canada. Parts of this work were presented at the 2017 IEEE International Workshop on Information Theory \cite{jjw2017}, the 2019 IEEE International Symposium on Information Theory \cite{jjw2019isit}, and the 2020 IEEE International Symposium on Information Theory \cite{jjw2020isit}.}
}

\maketitle

\begin{abstract}
   We propose an adaptive lossy joint source-channel coding (JSCC) scheme for sending correlated sources over two-terminal discrete-memoryless two-way channels (DM-TWCs). 
   The main idea is to couple the independent operations of the terminals via an adaptive coding mechanism, which can mitigate cross-interference resulting from simultaneous channel transmissions and concurrently exploit the sources' correlation to reduce the end-to-end reconstruction distortions. 
   Our adaptive JSCC scheme not only subsumes existing lossy coding methods for two-way simultaneous communication but also improves their performance.
   Furthermore, we derive outer bounds for our two-way lossy transmission problem and establish complete JSCC theorems in some special settings. 
   In these special cases, a non-adaptive separate source-channel coding (SSCC) scheme achieves the optimal performance, thus simplifying the design of the source-channel communication system.   
\end{abstract}


\section{Introduction}\label{sec:1}
Shannon's two-way communication \cite{shannon1961} considers full-duplex data transmission between two terminals. The terminals can send and receive data simultaneously on a shared two-way channel (TWC) without multiplexing \cite{jama2005} to make the best utilization of channel resources. 
The TWC was recently used as a building block in the construction of high spectral-efficiency transmission systems \cite{al2015, asadi2014, yuan2016}. However, designing an adaptive coding scheme for simultaneous transmission on TWCs remains challenging. More precisely, adaptive coding generates current channel inputs by taking into consideration past received signals. This mechanism conceptually improves the system's performance, but finding optimal coding methods remains elusive. 

\begin{figure}[!t]
  \centering
  \includegraphics[draft=false, scale=0.62]{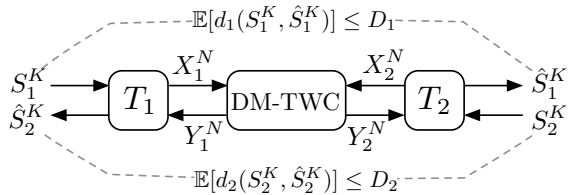}
  \caption{The block diagram for the lossy transmission of correlated source pair $(S_1^K, S_2^K)$ via $N$ uses of a noisy DM-TWC.}
  \label{fig:TWCblcok}
\end{figure}

In this paper, we investigate the adaptive coding problem from an information-theoretic perspective. 
Specifically, we consider the two-way lossy source-channel communication system depicted in Fig \ref{fig:TWCblcok}, where two terminals exchange correlated sources on a discrete-memoryless TWC (DM-TWC). 
Our objective is to characterize the achievable distortion region of the system for any given correlated sources, memoryless channel, transmission rate, and distortion measures. 
Before presenting our contributions, we first review existing results on two-way channel coding and source coding.

\subsection{Literature Review}\label{sec:review}
The capacity problem for general DM-TWCs is not yet completely solved in single-letter form.
In \cite{shannon1961}, Shannon presented a random coding inner bound and a cut-set outer bound to the capacity region. 
He also exploited channel symmetry properties \cite[Section~11]{shannon1961} to determine the capacity region in some special cases, which are further generalized in \cite{hekstra1989, varshney2013, chaaban2017, seo2019, jjw2019,jjw2019cwit}. 
For DM-TWCs with symmetry properties, it was shown that Shannon's inner bound is tight, and hence adaptive coding is not needed to achieve capacity. 
In the literature, there are other improved inner bounds \cite{han1984,schalkwijk1982,schalkwijk1983, kramer1998,sabag2018} and outer bounds \cite{zhang1986,hekstra1989}.
A common idea to improve on Shannon's inner bound is to coordinate the terminals' transmission via a stationary process. 
Although the terminals operate independently, the adaptive encoding procedure driven by the stationary process ultimately coordinates their encoding operations, thus jointly optimizing their transmissions. 
In the improved outer bounds, one typically seeks extra dependency among channel inputs. 

In two-terminal two-way lossy source coding, the DM-TWC in Fig.~\ref{fig:TWCblcok} is assumed to be noiseless.
In \cite{kaspi1985}, Kaspi established a rate-distortion (RD) region for this system,\footnote{Kaspi's original proof relies on tree codes using an intricate approach. A simper proof can be found in \cite[Section 20.3.3]{kim2011} based on the Wyzer-Ziv source coding scheme \cite{wyner1976}.} which characterizes the trade-off between source compression rate and distortion, under an interactive communication protocol. 
Specifically, the protocol divides the entire transmission period into small segments, and only one terminal sends data at each segment. 
With this protocol, each terminal can decode a coarse description of the other terminal's messages after observing a new segment of channel outputs. 
All decoded coarse descriptions are then treated as side-information to compress source messages until final reconstructions are obtained. 
In \cite{maor2006}, Maor and Merhav extended Kaspi's result within the application of successive source refinement. 
Another related two-way source coding problem, where each terminal is only interested in extracting hidden information related to the source messages of the other terminal, is tackled in \cite{vera2019} under the so-called collaborative information bottleneck problem. The rate-relevance trade-off is determined under Kaspi's transmission protocol. 

In addition to the above results, there are other extensions of the source coding problem such as two-way source coding with a helper \cite{permuter2010}, two-way multi-terminal source coding \cite{chia2011, vega2016}, and two-way function computation \cite{ma2011, shin2019}. 
The capacity problem was also studied for TWCs with memory \cite{jjw2019} and in a multi-terminal setting with more than two terminals such as multi-access/broadcast, Z, and interference TWCs \cite{cheng2014} and three-way channels \cite{ong2012, ong2013, chaaban2016}. 
These architectures are beyond the scope of this paper. 

\subsection{Notation and Problem Setup} 
We next introduce the notation used in the paper. 
The symbols $\mathbb{Z}_+$ and $\mathbb{R}_{\ge 0}$ denote the sets of positive integers and non-negative real numbers, respectively. 
For any $i\ge 1$, let $A^i\triangleq (A_1, A_2, \dots, A_i)$ denote a length-$i$ sequence of random variables with common alphabet $\mathcal{A}$. 
The realization of $A^i$ will be denoted by $a^{i}=(a_1, a_2, \dots, a_i)\in\mathcal{A}^i$, where $\mathcal{A}^i$ is the $i$-fold Cartesian product of $\mathcal{A}$. 
When the length $i$ is clear from the context, we may write $\bm{A}$ and $\bm{a}$ in lieu of $A^i$ and $a^i$, respectively. 
Throughout the paper, all alphabets are finite, except for the Gaussian case briefly considered in Section \ref{subsec:examples}. 
Moreover, we delineate each terminal by index $j$ or $j'$, where $j, j'\in\{1, 2\}$.
To simplify the presentation, we assume that $j\neq j'$ when these indices appear together. 
Furthermore, the $k$th source message of terminal~$j$ is denoted by $S_{j, k}$, and its reconstruction at terminal~$j'$ is given by $\hat{S}_{j, k}$; also, the $n$th channel input and output of terminal~$j$ are denoted by $X_{j, n}$ and $Y_{j, n}$, respectively. 
For these system variables, we use $\mathcal{S}_j$, $\hat{\mathcal{S}}_j$, $\mathcal{X}_j$, and $\mathcal{Y}_j$ to denote their respective alphabets. 
The standard notation $\mathbb{E}$ stands for the expectation operator and $\mathbbm{1}\{\cdot\}$ stands for the indicator function.

We are now ready to define our problem. As depicted in Fig.~\ref{fig:TWCblcok}, two terminals exchange a block of correlated source messages $(S_1^K, S_2^K)$ of length-$K$ via $N$ uses of a noisy TWC. 
Terminal~$j$ only observes $S_j^K$ and intends to reconstruct $S^K_{j'}$ from $S_j^K$ and $Y_j^N$ subject to a distortion constraint.
Here, we assume that the source pairs $(S_{1, k}, S_{2, k})$, $1\le k\le K$, are independent and have the common joint probability distribution $P_{S_1, S_2}$; i.e., $P_{S_1^K, S_2^K}(s_1^K, s_2^K)=\prod_{k=1}^K P_{S_1, S_2}(s_{1, k}, s_{2, k})$, where $(s_{1, k}, s_{2, k})\in\mathcal{S}_1\times\mathcal{S}_2$. 
The distortion for the reconstruction $\hat{s}_j^K$ of source message $s_j^K$ is assessed via $d_j(s_j^K, \hat{s}_j^K)\,\triangleq\,K^{-1}\sum_{k=1}^K d_j({s_{j, k}, \hat{s}_{j, k}})$, where $d_j: \mathcal{S}_j\times\mathcal{\hat{S}}_j{\rightarrow}\mathbb{R}_{\ge 0}$ is a single-letter distortion measure for source $S_j$.  
Furthermore, the noisy TWC is used without adopting any interactive communication protocol such as in \cite{kaspi1985, maor2006}. 
We only consider DM-TWCs with input alphabet $\mathcal{X}_j$ and output alphabet $\mathcal{Y}_j$ for terminal~$j$, $j=1, 2$, and with transition probability $P_{Y_1, Y_2|X_1, X_2}$.
More precisely, we have that $P_{Y_{1, n}, Y_{2, n}|X_1^n, X_2^n, Y_1^{n-1}, Y_2^{n-1}}=P_{Y_{1, n}, Y_{2, n}|X_{1, n}, X_{2, n}}=P_{Y_1, Y_2|X_1, X_2}$ 
for all $n$. 
For this system setup, we seek forward and converse coding theorems for lossy source-channel transmissibility. 

\subsection{Related Work and Our Approach}
To the best of our knowledge, there are only few works related to our problem setup. 
In \cite[Section 14]{shannon1961}, Shannon implicitly illustrated that perfect matching among the source and channel statistics and alphabets results in error-free communication, with the optimal scheme given by uncoded transmission. 
In \cite{maor2006}, the JSCC problem was studied for DM-TWCs which consist of two independent one-way channels. 
Together with the protocol mentioned in Section~\ref{sec:review}, Kaspi's source coding result was extended for successive source refinement.
Also, a complete JSCC theorem was derived in this particular setting.      
By contrast, the authors in \cite[Section VIII]{gunduz2009} tackled the two-way transmission problem for general DM-TWCs without deploying any protocol.
The correlation-preserving coding scheme of \cite{cover1980} was adopted for almost lossless transmission; i.e., when requiring the block error rate of the source reconstructions to vanish asymptotically. 
Similar to Shannon's idea, the (non-adaptive) coding scheme of \cite{gunduz2009} can preserve source correlation in the channel inputs to facilitate two-way transmission; however, it does not apply to the lossy setup. 
In this paper, we tackle a transmission problem that is more general in many aspects; e.g., we do not consider a particular type of DM-TWC or assume a given communication protocol. 
We next sketch the concepts behind our main JSCC achievability result. 

As the transmissions of the terminals influence each other on a shared channel and generally cause cross-interference, we propose to design the coding strategies jointly. 
For this purpose, we construct joint source-channel codes that induce a stationary Markov chain that couples all variables of the communication system in Fig.~\ref{fig:TWCblcok}. 
In principle, when the channel inputs are generated by such codes, all system variables will behave according to the stationary distribution of the induced chain, thus coordinating the independent transmissions of the terminals. 
Specifically, we combine the following coding techniques to build our adaptive codes. 
First, we adopt the functional form of superposition coding \cite{cover1972} to generate channel inputs, which plays a central role in inducing the desired Markov transmission process.
We also modify the analog/digital hybrid coding scheme of \cite{kim2015} to exploit side-information for decoding, in addition to its original source-correlation-preserving mechanism. 
Moreover, we use past channel inputs and outputs similarly to \cite{han1984} to enable adaptive coding. 
We note that although these techniques are not new, combining and integrating them into an adaptive two-way coding framework for our problem setup is challenging. 
We next summarize the contributions of the paper. 


\subsection{Summary of Contributions}
Our primary contribution is the construction of an adaptive coding scheme to prove a forward JSCC theorem; but we also derive some converse results and complete JSCC theorems. 
The details are as follows.

\noindent$\bullet$ \textbf{Inner Bounds and Examples}: a general JSCC result (Theorem~\ref{thm:main}) for two-way lossy simultaneous transmission is established using the concepts of hybrid analog/digital coding, superposition coding, and adaptive channel coding, together with a low-complexity sliding-window decoder. Two simplified achievability results (Corollaries~\ref{cor:twchybrid} and~\ref{cor:ssccwzhan}) are derived from the main theorem. 
Moreover, our coding method is shown to subsume some basic schemes such as uncoded transmission and the concatenation of Wyner-Ziv (WZ) source coding and Shannon's (or Han's) channel coding; it also recovers the almost lossless transmission of \cite{gunduz2009}. 
Four illustrated examples (Examples 1-4) are provided to highlight the difference between the coding schemes. 

\noindent$\bullet$ \textbf{Outer Bounds and Complete JSCC Theorems}: two outer bounds (Lemmas~\ref{lma:OB1} and \ref{lma:OB2}) to the achievable distortion region are obtained using standard arguments. 
The bounds are expressed in terms of the standard RD function and the conditional RD function and are hence easy to compute for many classical models of correlated sources. 
Furthermore, four complete theorems (Theorems~\ref{thm:JSCC1}-\ref{thm:JSCC4}) that fully characterize the achievable distortion region for certain system settings are obtained. 
Specifically, for DM-TWCs with symmetry properties \cite{jjw2019}, we show the optimality of SSCC in the following settings: 
\begin{itemize}
  \item lossy transmission of independent sources; 
  \item almost lossless transmission of correlated sources;
  \item lossy transmission of correlated sources whose WZ and conditional RD functions are equal;
  \item lossy transmission of correlated sources having a common part in the sense of G\'acs-K\"orner-Witsenhausen \cite[Section 14.2.2]{kim2011}.
\end{itemize}
Examples for Theorems~\ref{thm:JSCC3} and~\ref{thm:JSCC4} (Examples~5-7) are also provided. 

The rest of the paper is organized as follows. 
In Section~\ref{sec:Preliminary}, definitions and background information are provided. 
Our forward coding theorem is presented in Section~\ref{sec:main}; its full proof is provided in Appendices~\ref{subsec:mainproof} and~\ref{subsec:claims}. 
Simplified versions of the main theorem are given in Section~\ref{sec:speicalcases}, together with a derivation of the associated coding schemes.
Section~\ref{sec:conandjssc} establishes converse results and complete JSCC theorems. 
Examples and a discussion are given in Section \ref{sec:exdiss}, and conclusions are drawn in Section \ref{sec:conclusion}.

\section{Preliminaries}\label{sec:Preliminary}
In this section, we define joint source-channel codes and the achievable distortion region for source-channel communication over a TWC. We also review various RD function expressions for point-to-point communication and channel coding results for DM-TWCs, which will be used in Section~\ref{sec:speicalcases}. 

\subsection{Definitions}\label{sec:definitions}
For our problem setup, a joint source-channel code is defined as follows. 

\begin{definition}
An $(N, K)$ code for transmitting $(S_1^K, S_2^K)$ over a DM-TWC consists of two sequences of encoding functions $\underline{f}_1\triangleq \{f_{1, n}\}_{n=1}^N$ and $\underline{f}_2\triangleq \{f_{2, n}\}_{n=1}^N$ such that 
\[
\begin{array}{ll}
f_{1,1}: \mathcal{S}_1^K\to\mathcal{X}_1,& f_{1, n}: \mathcal{S}_1^K\times\mathcal{Y}_1^{n-1}\to\mathcal{X}_1\\
f_{2,1}: \mathcal{S}_2^K\to\mathcal{X}_2,& f_{2, n}: \mathcal{S}_2^K\times\mathcal{Y}_2^{n-1}\to\mathcal{X}_2
\end{array}
\]
for $n=2, 3, \dots, N$, and two decoding functions $g_1:\mathcal{S}_1^K\times\mathcal{Y}_1^N\to\hat{\mathcal{S}}_2^K$ and $g_2:\mathcal{S}_2^K\times\mathcal{Y}_2^N\to\hat{\mathcal{S}}_1^K$.
\end{definition}

The channel inputs at time $n=1$ are only functions of the source messages, i.e., $X_{j, 1} = f_{j, 1}(S_j^K)$, but the subsequent channel inputs are generated by also adapting to the previous channel outputs via $X_{j, n} = f_{j, n}(S_j^K, Y_j^{n-1})$ for $n=2, 3, \dots, N$. 
Such encoding strategy is called adaptive coding, in contrast to its non-adaptive counterpart where $X_{j, n}=f_{j, n}(S_j^k)$ for all~$n$. 
We remark that our code definition also involves block-wise decoding; i.e., terminal $j$ reconstructs $S^K_{j'}$ via $\hat{S}^K_{j'}=g_j(S_j^K, Y_j^N)$ after receiving the entire $N$ channel outputs. 

Moreover, the rate of the joint source-channel code is given by $K/N$ (source symbols/channel use), and the associated expected distortion is $D_j(K)\triangleq\mathbb{E}[d_j(S_j^K, \hat{S}_j^K)]$, where the expectation is taken with respect to the joint probability distribution
\begin{IEEEeqnarray}{l}
  \scalemath{0.95}{P_{S_1^K, S_2^K, X_1^N, X_2^N, Y_1^N, Y_2^N}=P_{S_1^K, S_2^K}\Bigg(\prod\limits_{n=1}^N P_{X_{1, n}|S_1^K, Y_1^{n-1}}\Bigg)}\scalemath{0.95}{\Bigg(\prod\limits_{n=1}^N P_{X_{2, n}|S_2^K, Y_2^{n-1}}\Bigg)\Bigg(\prod\limits_{n=1}^N P_{Y_{1, n}, Y_{2, n}|X_{1, n}, X_{2, n}}\Bigg)},\nonumber
\end{IEEEeqnarray}
where $P_{Y_{1, n}Y_{2,n}|X_{1, n},X_{2, n}}=P_{Y_1,Y_2|X_1,X_2}$ for $n=1, 2, \dots,N$ (determined by the DM-TWC).

\begin{definition}
A distortion pair $(D_1, D_2)$ is said to be achievable at rate $R$ if there exists a sequence of $(N, K)$ joint source-channel codes (where $N$ is a function of $K$) such that $\lim_{K\to\infty} K/N = R$ and $\limsup_{K\to\infty}\allowbreak D_j(K)\le D_j$, $j=1, 2$. 
The achievable distortion region of a rate-$R$ two-way lossy transmission system is the convex closure of the set of all achievable distortion pairs (at rate $R$). 
\end{definition}\smallskip




\subsection{Rate-Distortion Functions}\label{subsec:RDFs}
As a DM-TWC can be viewed as two state-dependent one-way channels, the following source coding related functions (each expressed in terms of a constrained minimization of a mutual information quantity) for one-way systems are also useful in the two-way channel setup.  
\begin{itemize}
  \item Standard RD function \cite[Sec. 3.6]{kim2011}: 
  \begin{IEEEeqnarray}{l}
    R^{(j)}(D_j)=\min\limits_{P_{\hat{S}_j|S_j}:\mathbb{E}[d_j(S_j, \hat{S}_j)]\le D_j} I(S_j; \hat{S}_j).\label{eq:RD}   
  \end{IEEEeqnarray}
  \smallskip
  \item WZ-RD function \cite{wyner1976}: Letting $T_j\in\mathcal{T}_j$ with $|\mathcal{T}_j|\le |\mathcal{S}_j|+1$ denote an auxiliary random variable that satisfies the Markov chain $T_j\markov S_j\markov S_{j'}$, we have
  \begin{IEEEeqnarray}{l}
    R^{(j)}_{\text{WZ}}(D_j)=\min\limits_{P_{T_j|S_j}}\min_{\substack{h: \mathcal{T}_{j}\times\mathcal{S}_{j'}\rightarrow\hat{\mathcal{S}}_j\\ \mathbb{E}[d_j(S_j, h(T_j, S_{j'})]\le D_j}} I(S_j; T_j|S_{j'}).\label{eq:WZRD}
  \end{IEEEeqnarray}
  \smallskip
  \item  Conditional RD function \cite{gray1972}: 
  \begin{IEEEeqnarray}{l}
    R_{S_j|S_{j'}}(D_j)=\min_{\substack{P_{\hat{S}_j|S_1, S_2}\\ \mathbb{E}[d_j(S_j, \hat{S}_j)]\le D_j}} I(S_j; \hat{S}_j|S_{j'}).\label{eq:condRD}    
  \end{IEEEeqnarray}
\end{itemize}

We remark that the source coding schemes that achieve the standard RD and WZ-RD functions can be the building blocks of an SSCC scheme for our overall system. 
For example, terminal~$j$ can apply the WZ coding scheme to compress source $S_j^K$ given side-information $S_{j'}^K$. 
Although the coding scheme that achieves the conditional RD function cannot be applied in our problem setup  (since there is no common side-information at the encoder and the decoder in general), the scheme is useful when $S_1$ and $S_2$ have a common part in the sense of G\'acs-K\"orner-Witsenhausen \cite[Section 14.2.2]{kim2011}.
We will use this result in Theorem~\ref{thm:JSCC4} (see Section~\ref{sec:JSCCthms}). 

\subsection{Capacity Bounds for DM-TWCs}\label{subsec:CBs}
To introduce capacity bounds for DM-TWCs, we first give some definitions. 
Roughly speaking, an $(N, R_{\text{c}, 1}, R_{\text{c}, 2})$ channel code for a DM-TWC is defined similarly to an $(N, K)$ joint source-channel code, except that the correlated sources $S_1^K$ and $S_2^K$ are replaced with  independent and uniformly distributed random indices $I_1\in\mathcal{I}_1$ and $I_2\in\mathcal{I}_2$, respectively, where $|\mathcal{I}_1|=2^{\mli{NR}_{\text{c}, 1}}$ and $|\mathcal{I}_2|=2^{\mli{NR}_{\text{c}, 2}}$.  
As a result, two-way channel codes can incorporate or exclude adaptive coding.
Given a DM-TWC, a channel coding rate pair $(R_{\text{c}, 1}, R_{\text{c}, 2})$ is called achievable if there exists a sequence of $(N, R_{\text{c}, 1}, R_{\text{c}, 2})$ channel codes such that $I_1$ and $I_2$ can be reliably exchanged (i.e., with asymptotically vanishing decoding error probability). 
The capacity region is defined as the convex closure of the set of all achievable rate pairs.

To date, a single-letter characterization of the capacity region of general DM-TWCs is still not found. 
In \cite{shannon1961}, Shannon derived the inner bound region
\begin{IEEEeqnarray}{l}
  \overline{\text{co}}\left( \bigcup_{P_{X_1}P_{X_2}}\big\{(R_{\text{c}, 1}, R_{\text{c}, 2}): R_{\text{c}, 1} <  I(X_1; Y_2|X_2), R_{\text{c}, 2} < I(X_2; Y_1|X_1\big\}\right)\label{eq:shannonIB}
\end{IEEEeqnarray}
and the outer bound region
\begin{IEEEeqnarray}{l}
  \bigcup_{P_{X_1, X_2}}\big\{(R_{\text{c}, 1}, R_{\text{c}, 2}): R_{\text{c}, 1} <  I(X_1; Y_2|X_2), R_{\text{c}, 2} < I(X_2; Y_1|X_1)\big\}\nonumber
\end{IEEEeqnarray} 
for channel capacity, where $\overline{\text{co}}(\cdot)$ denotes taking the closure of the convex hull.
In general, the two capacity bounds do not coincide, but they match each other for channels with symmetry properties; i.e., DM-TWCs that satisfy the channel symmetry conditions in either \cite[Theorem~1]{jjw2019} or \cite[Theorem~4]{jjw2019}. 
For these ``symmetric'' DM-TWCs, the capacity region can be exactly determined via non-adaptive coding and is given by the set of all achievable rate pairs in \eqref{eq:shannonIB} under independent inputs.
Moreover, taking the convex closure in \eqref{eq:shannonIB} is not needed.

Shannon's inner bound result was later improved by Han \cite{han1984} under an adaptive channel coding scheme, showing that any rate pair in the following region is achievable:
\begin{IEEEeqnarray}{l}
  \overline{\text{co}}\left( \bigcup_{P_{\tilde{V}_1, \tilde{V}_2, \tilde{W}_1, \tilde{W}_2, X_1, X_2}}\big\{(R_{\text{c}, 1}, R_{\text{c}, 2}): R_{\text{c}, 1} <  I(\tilde{V}_1; X_2, Y_2, \tilde{V}_2, \tilde{W}_2),
  R_{\text{c}, 2} < I(\tilde{V}_2; X_1, Y_1, \tilde{V}_1, \tilde{W}_1)\big\}\right)\nonumber
\end{IEEEeqnarray}
where the joint probability distribution $P_{\tilde{V}_1, \tilde{V}_2, \tilde{W}_1, \tilde{W}_2, X_1, X_2}$ is defined in \cite[Section IV]{han1984}.\footnote{The random variables $\tilde{V}_j$ and $\tilde{W}_j$ correspond to the random variables $\tilde{U}_j$ and $\tilde{W}_j$ in Han's  scheme, respectively.} 
We note that Kramer further generalized Han's result from a concatenated coding perspective \cite[Section 4.3.2]{kramer1998} with achievable rate pairs obtained in terms of conditional directed mutual information quantities using a random coding error exponent analysis under maximum-likelihood decoding \cite{gallager1968}. 
In this paper, as we pursue single-letter expressions, we mainly focus on Shannon's and Han's results. 


\section{Forward JSCC Theorem Based on Adaptive Coding}\label{sec:main}
This section establishes the most general achievability result in the paper.  
Without loss of generality, we only consider rate-one transmission, i.e., $N=K$; other rates can be obtained via suitable super-symbols.\footnote{To obtain a rate-$\frac{K_1}{N_1}$ result, we define a super source symbol (resp., a super channel input/output symbol) by combining $K_1$ source symbols (resp., $N_1$ channel input/output symbols).}
First of all, we describe the key technical ingredients used in obtaining the main result in Theorem~\ref{thm:main}.
Our approach is to construct an extended channel (from the original DM-TWC) and use a stationary Markov chain to coordinate the terminals' transmissions.

%

\subsection{Two-Way Coded Channel}\label{sec:codedTWC}
\begin{figure}[t!]
  \centering
  \includegraphics[draft=false, scale=0.6]{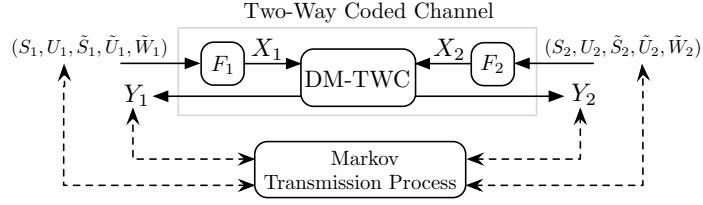}
  \caption{An illustration of the transmission over the two-way coded channel.}
  \label{fig:codedtwc}
\end{figure}

Consider an auxiliary coded channel built on the original (physical) DM-TWC, as shown in the central box of Fig.~\ref{fig:codedtwc}. 
The coded channel has inputs $S_j, U_j, \tilde{S}_j, \tilde{U}_j$ and $\tilde{W}_j$ at terminal~$j$. 
The input pairs $(S_j, U_j)$ and $(\tilde{S}_j, \tilde{U}_j)$ are used to carry the current and some prior source information, respectively, where $U_j$ (resp., $\tilde{U}_j$) denotes the coded version of $S_j$ (resp., $\tilde{S}_j$). 
The input $\tilde{W}_j$ carries some past channel inputs and outputs at terminal~$j$. 
The new channel also involves two encoding functions $F_j: \mathcal{S}_j\times \mathcal{U}_j\times\tilde{\mathcal{S}}_j\times\tilde{\mathcal{U}}_j\times\tilde{\mathcal{W}}_j\to\mathcal{X}_j$, which transform the inputs of the coded channel into the inputs for the original DM-TWC. 
The outputs of the new channel are still $Y_1$ and $Y_2$. 
The joint input probability distribution of the coded channel is given by
\[P_{S_1, S_2, U_1, U_2, \tilde{S}_1, \tilde{S}_2, \tilde{U}_1, \tilde{U}_2, \tilde{W}_1, \tilde{W}_2}= P_{S_1, S_2}P_{U_1|S_1}P_{U_2|S_2}P_{\tilde{S}_1, \tilde{S}_2, \tilde{U}_1, \tilde{U}_2, \tilde{W}_1, \tilde{W}_2},\] 
and the transition probability of the coded channel is given by 
\begin{IEEEeqnarray}{l}
  P_{Y_1, Y_2|S_1, S_2, U_1, U_2, \tilde{S}_1, \tilde{S}_2, \tilde{U}_1, \tilde{U}_2, \tilde{W}_1, \tilde{W}_2}(y_1, y_2|s_1, s_2, u_1, u_2, \tilde{s}_1, \tilde{s}_2, \tilde{u}_1, \tilde{u}_2, \tilde{w}_1, \tilde{w}_2)\nonumber\\
  \ \ \ =\sum_{x_1, x_2} \mathbbm{1}\{x_1=F_1(s_1, u_1, \tilde{s}_1, \tilde{u}_1, \tilde{w}_1)\}\mathbbm{1}\{x_2=F_2(s_2, u_2, \tilde{s}_2, \tilde{u}_2, \tilde{w}_2)\}P_{Y_1, Y_2|X_1, X_2}(y_1, y_2|x_1, x_2).\label{eq:sysprob}\nonumber\\*\IEEEeqnarraynumspace
\end{IEEEeqnarray}

\vspace{-0.6cm}
\subsection{Markov Chain for the Coded Channel}\label{sec:markovforcodedTWC}
For the repeated use over time of the two-way coded channel, we next construct a discrete-time Markov chain for the overall system with state space: \[\mathcal{S}_1\times\mathcal{S}_2\times \mathcal{U}_1\times \mathcal{U}_2\times \tilde{\mathcal{S}}_1\times \tilde{\mathcal{S}}_2\times \tilde{\mathcal{U}}_1\times \mathcal{\tilde{U}}_2\times\tilde{\mathcal{W}}_1\times \tilde{\mathcal{W}}_2\times\mathcal{X}_1\times\mathcal{X}_2\times\mathcal{Y}_1\times\mathcal{Y}_2,\] 
where $\tilde{\mathcal{S}}_j\triangleq\mathcal{S}_j$, $\tilde{\mathcal{U}}_j\triangleq\mathcal{U}_j$, and $\tilde{\mathcal{W}}_j\triangleq\mathcal{X}_j\times\mathcal{Y}_j$ for $j=1, 2$. 
This Markov chain will be used to coordinate the transmissions of the two terminals as shown in Fig.~\ref{fig:codedtwc}. 
Let 
\[Z^{(t)}\triangleq(S^{(t)}_1, \allowbreak S^{(t)}_2, \allowbreak U^{(t)}_1, \allowbreak U^{(t)}_2, \allowbreak\tilde{S}^{(t)}_1, \allowbreak\tilde{S}^{(t)}_2, \allowbreak\tilde{U}^{(t)}_1, \allowbreak\tilde{U}^{(t)}_2, \allowbreak\tilde{W}^{(t)}_1, \allowbreak\tilde{W}^{(t)}_2, X^{(t)}_1, \allowbreak X^{(t)}_2, \allowbreak Y^{(t)}_1, \allowbreak Y^{(t)}_2)\]
denote the state of the Markov chain at time $t\in\mathbb{Z}_{+}$, where we set $\tilde{S}^{(t)}_j\triangleq S^{(t-1)}_j$, $\tilde{U}^{(t)}_j\triangleq U^{(t-1)}_j$, and $\tilde{W}^{(t)}_j\triangleq \allowbreak(X^{(t-1)}_j, Y^{(t-1)}_j)$. 
Given a parameter tuple $(P_{U_1|S_1}, P_{U_2|S_2}, P_{\tilde{S}_1, \tilde{S}_2, \tilde{U}_1, \tilde{U}_2, \tilde{W}_1, \tilde{W}_2}, \allowbreak  F_1, \allowbreak F_2)$, we generate the quadruple $(S^{(t)}_1, S^{(t)}_2, U^{(t)}_1, U^{(t)}_2)$ for all $t$ according to $P_{S_1, S_2, U_1, U_2}= \allowbreak P_{S_1, S_2}\allowbreak  P_{U_1|S_1}\allowbreak P_{U_2|S_2}$ independently of $(\tilde{S}^{(t)}_1, \allowbreak\tilde{S}^{(t)}_2, \allowbreak\tilde{U}^{(t)}_1, \allowbreak\tilde{U}^{(t)}_2, \allowbreak\tilde{W}^{(t)}_1, \allowbreak\tilde{W}^{(t)}_2)$.  
The tuple $(\tilde{S}^{(1)}_1, \allowbreak\tilde{S}^{(1)}_2, \allowbreak\tilde{U}^{(1)}_1, \allowbreak\tilde{U}^{(1)}_2, \allowbreak\tilde{W}^{(1)}_1, \allowbreak\tilde{W}^{(1)}_2)$ is initialized according to $P_{\tilde{S}_1, \tilde{S}_2, \tilde{U}_1, \tilde{U}_2, \tilde{W}_1, \tilde{W}_2}$. 
The physical channel input at terminal~$j$ is naturally produced as $X^{(t)}_j=F_j(S_j^{(t)}, U_j^{(t)}, \tilde{S}_j^{(t)}, \tilde{U}_j^{(t)},\allowbreak \tilde{W}_j^{(t)})$, and the received channel output is $Y^{(t)}_j$. 
Based on this construction, the transition kernel of $\{Z^{(t)}\}$ is given by  
\begin{IEEEeqnarray}{l}
P_{Z^{(t)}|Z^{(t-1)}}(\scalemath{0.9}{s_1, s_2, u_1, u_2, \tilde{s}_1, \tilde{s}_2, \tilde{u}_1, \tilde{u}_2, \tilde{w}_1, \tilde{w}_2, x_1, x_2, y_1, y_2|s'_1, s'_2, u'_1, u'_2, \tilde{s}'_1, \tilde{s}'_2, \tilde{u}'_1, \tilde{u}'_2, \tilde{w}'_1, \tilde{w}'_2, x'_1, x'_2, y'_1, y'_2})\nonumber\\
    \ = P_{S_1, S_2}(s_1, s_2)P_{U_1|S_1}(u_1|s_1)P_{U_2|S_2}(u_2|s_2)\mathbbm{1}\{\tilde{s}_1=s'_1\}\mathbbm{1}\{\tilde{s}_2=s'_2\}\mathbbm{1}\{\tilde{u}_1=u'_1\}\mathbbm{1}\{\tilde{u}_2=u'_2\}\nonumber\\
    \qquad\qquad\qquad\cdot\mathbbm{1}\{\tilde{w}_1=(x'_1, y'_1)\}\mathbbm{1}\{\tilde{w}_2=(x'_2, y'_2)\}\mathbbm{1}\{x_1=F_1(s_1, u_1, \tilde{s}_1, \tilde{u}_1, \tilde{w}_1)\}\nonumber\\
    \qquad\qquad\qquad\qquad\qquad\qquad\cdot\mathbbm{1}\{x_2=F_2(s_2, u_2, \tilde{s}_2, \tilde{u}_2, \tilde{w}_2)\}P_{Y_1, Y_2|X_1, X_2}(y_1, y_2|x_1, x_2)\label{eq:trankernel}\IEEEeqnarraynumspace
\end{IEEEeqnarray}
for $t\ge 2$. 
It is easy to see that the process $\{Z^{(t)}\}$ is a first-order time-homogeneous Markov chain. 
However, whether or not the chain is stationary depends on the given parameters. 

\subsection{Stationary Distribution under Distortion Constraints}\label{sec:stationarydist}
To obtain an achievability result with time-independent conditions, we only consider a stationary Markov chain. 
The following procedure can be used to find its parameters. 
Given $P_{S_1, S_2}$ and $P_{Y_1, Y_2|X_1, X_2}$, we first fix a choice of $P_{U_j|S_j}$ and $F_j$, $j=1, 2$, and write the transition kernel \eqref{eq:trankernel} in matrix form as $Q_Z$. 
The matrix $Q_Z$ is stochastic, and since all alphabets are finite, an eigenvector of $Q_Z$ associated with the eigenvalue $1$ exists and gives a stationary distribution $P_{Z}$ for $\{Z^{(t)}\}$, i.e., $P_{Z}=P_{Z}Q_Z$. 
Clearly, using the marginal distribution $P_{\tilde{S}_1, \tilde{S}_2, \tilde{U}_1, \tilde{U}_2, \tilde{W}_1, \tilde{W}_2}$ of $P_{Z}$ with the chosen $P_{U_j|S_j}$ and $F_j$, $j=1, 2$, to initialize the Markov chain ensures stationarity. 
Note that for the stationary chain the two independent quadruples $(S^{(t)}_1, S^{(t)}_2, U^{(t)}_1, U^{(t)}_2)$ and $(\tilde{S}^{(t)}_1, \tilde{S}^{(t)}_2, \allowbreak\tilde{U}^{(t)}_1, \allowbreak\tilde{U}^{(t)}_2)$ have identical distributions for all $t$; thus $P_{S_1, S_2, U_1, U_2}=P_{\tilde{S}_1, \tilde{S}_2, \tilde{U}_1, \tilde{U}_2}$. 
Moreover, due to our construction of $\{Z^{(t)}\}$, we have the following necessary conditions for stationarity
\begin{IEEEeqnarray}{l}
  P_{S_1, S_2}=P_{\tilde{S}_1, \tilde{S}_2},\label{eq:stationary1}\\
  P_{U_j|S_j}=P_{\tilde{U}_j|\tilde{S}_j},\label{eq:stationary2}
\end{IEEEeqnarray}
for $j=1, 2$. For source reconstruction, we next associate the parameters with decoding functions\footnote{As will be seen at the end of the section or in Appendix~\ref{subsec:mainproof}, terminal~$j$ reconstructs the prior source message $\tilde{S}_{j'}$ as $\hat{\tilde{S}}_{j'}$ after recovering $\tilde{U}_{j'}$; this reconstruction is done via $G_j$.} $G_j: \tilde{\mathcal{U}}_{j'}\times\mathcal{S}_j\times \mathcal{U}_j \times\tilde{\mathcal{S}}_j\times\tilde{\mathcal{U}}_j\times\tilde{\mathcal{W}}_j\times {\mathcal{Y}}_j\to\hat{\tilde{\mathcal{S}}}_{j'}$, $j=1, 2$. 
For simplicity, we call the tuple $(P_{U_1|S_1}, P_{U_2|S_2}, P_{\tilde{S}_1, \tilde{S}_2, \tilde{U}_1, \tilde{U}_2}, P_{\tilde{W}_1, \tilde{W}_2|\tilde{S}_1, \tilde{S}_2, \tilde{U}_1, \tilde{U}_2}, F_1, F_2, G_1, G_2)$ a {\it configuration}, which specifies a stationary distribution $P_Z$ given by 
\begin{IEEEeqnarray}{l}
P_{Z}=\underbrace{P_{S_1, S_2}P_{U_1|S_1}P_{U_2|S_2}}_{=P_{S_1, S_2, U_1, U_2}}\underbrace{P_{\tilde{S}_1, \tilde{S}_2}P_{\tilde{U}_1|\tilde{S}_1}P_{\tilde{U}_2|\tilde{S}_2}}_{=P_{\tilde{S}_1, \tilde{S}_2, \tilde{U}_1, \tilde{U}_2}}P_{\tilde{W}_1, \tilde{W}_2|\tilde{S}_1, \tilde{S}_2, \tilde{U}_1, \tilde{U}_2}\nonumber\\ 
\qquad\qquad\qquad\qquad \cdot P_{X_1|S_1, U_1, \tilde{S}_1, \tilde{U}_1, \tilde{W}_1}P_{X_2|S_2, U_2, \tilde{S}_2, \tilde{U}_2, \tilde{W}_2}P_{Y_1, Y_2|X_1, X_2},\nonumber   
\end{IEEEeqnarray}
where $P_{S_1, S_2}$ and $P_{Y_1, Y_2|X_1, X_2}$ are fixed by the problem setup and $P_{X_j|S_j, U_j, \tilde{S}_j, \tilde{U}_j, \tilde{W}_j}$ is determined by $F_j$, $j=1, 2$. 
We also let $\Pi_{Z}(D_1, D_2)$ denote the set of all configurations that induce a stationary chain and satisfy the distortion constraints: $\mathbb{E}[d_j(\tilde{S}_j, \hat{\tilde{S}}_j)]\le D_j$ for $j=1, 2$.

\subsection{Main Result: JSCC Achievability}\label{subsec:mainD}
Based on the above setup, we establish the achievability result in Theorem~\ref{thm:main} below. 
The full proof is provided in Appendices~\ref{subsec:mainproof} and~\ref{subsec:claims}.
In Theorem~\ref{thm:main}, one can further convexify the achievable distortion region via a standard time-sharing argument \cite{cover2006}.

\begin{theorem}[Adaptive JSCC]
  \label{thm:main}
  A distortion pair $(D_1, D_2)$ is achievable for the rate-one lossy transmission of correlated sources over a DM-TWC if there exists a configuration in $\Pi_{Z}(D_1, D_2)$ such that
  \begin{IEEEeqnarray}{rCl}
    \label{eq:mainconds}
    \IEEEyesnumber
    \IEEEyessubnumber*
    I(\tilde{S}_1; \tilde{U}_1)&< & I(\tilde{U}_1; S_2, U_2, \tilde{S}_2, \tilde{U}_2, \tilde{W}_2, X_2, Y_2),\label{eq:mainconda}\\
    I(\tilde{S}_2; \tilde{U}_2)&< & I(\tilde{U}_2; S_1, U_1, \tilde{S}_1, \tilde{U}_1, \tilde{W}_1, X_1, Y_1).\label{eq:maincondb}
  \end{IEEEeqnarray}
\end{theorem}

\begin{figure*}[!h]
  \centering
  \subfloat[The encoding process of terminal~$j$, where each node represents a block of variables and each node is a function of other nodes specified by the incoming edges.]{\includegraphics[draft=false, scale=0.61]{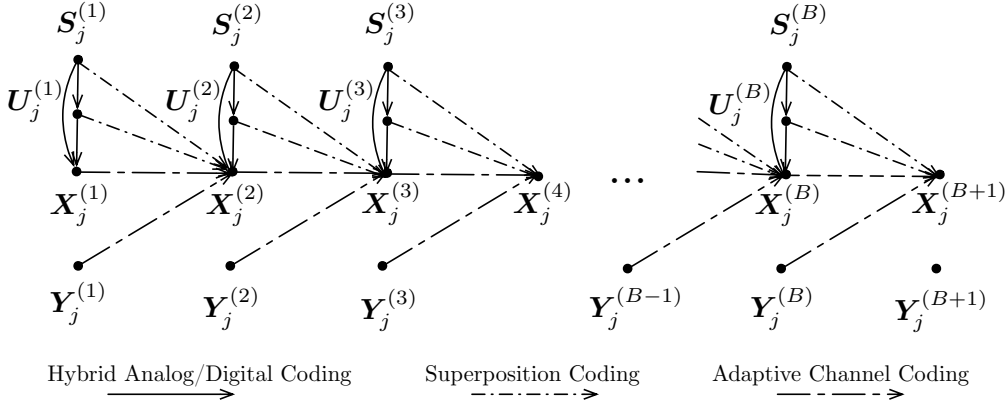}
  \label{fig:ENC}}
  \\
  \subfloat[The block diagram for sliding-window decoding.]{\includegraphics[draft=false, scale=0.58]{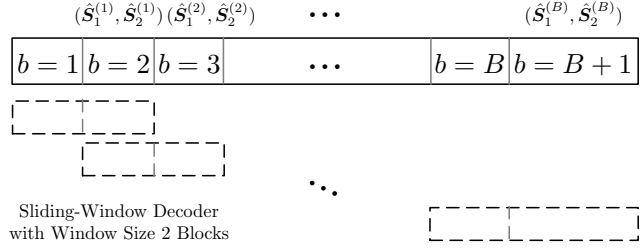}
  \label{fig:DEC}}
  \caption{An illustration of the proposed JSCC method.}
  \label{fig:ENCDEC}
\end{figure*}

To facilitate the understanding of the conditions in \eqref{eq:mainconds}, we sketch our coding method used in the proof, which extends the hybrid analog/digital coding scheme of \cite{kim2015}, used in conjunction with superposition block Markov encoding \cite{cover1979relay, han1984} and a sliding-window decoder, as shown in Fig.~\ref{fig:ENCDEC}.
In our method, instead of exchanging a single block of source message pairs $(S_1^K, S_2^K)$ via $K$ channel uses, we exchange $B$ blocks of such source message pairs via $K(B+1)$ channel uses for some $B\in\mathbb{Z}_{+}$.
The overall transmission rate is $\frac{B}{B+1}$, which approaches $1$ as $B$ goes to infinity.   
The extra $K$ channel uses can be viewed as added redundancy for data protection. 

For $1\le b\le B$, let $\bm{S}_j^{(b)}=(S_{j, 1}^{(b)}, S_{j, 2}^{(b)}, \dots, S_{j, K}^{(b)})$ denote the $b$th source message block at terminal~$j$; the same indexing convention applies to other variables.
As shown in Fig.~\ref{fig:ENC},\footnote{To simplify the presentation of our encoding scheme, we write $\bm{S}_j^{(b-1)}, \bm{U}_j^{(b-1)}$, and $(\bm{X}_j^{(b-1)}, \bm{Y}_j^{(b-1)})$ in lieu of $\tilde{\bm{S}}^{(b)}_j$, $\tilde{\bm{U}}^{(b)}_j$, and $\tilde{\bm{W}}^{(b)}_j$, respectively, to refer to the prior information variables at block instance $b$, for $2\le b\le B+1$. Later, when presenting our decoder, we also use $\hat{\bm{S}}_j^{(b-1)}$ (resp., $\hat{\bm{U}}_j^{(b-1)}$) rather than $\hat{\tilde{\bm{S}}}_j^{(b)}$ (resp., $\hat{\tilde{\bm{U}}}_j^{(b)}$) to denote the reconstruction of $\tilde{\bm{S}}_j^{(b)}$ (resp., $\tilde{\bm{U}}_j^{(b)}$).} the encoding involves hybrid analog/digital coding, superposition coding, and adaptive channel coding. 
In the $b$th transmission block, terminal~$j$ first encodes its source message $\bm{S}_j^{(b)}$ into the digital codeword $\bm{U}_j^{(b)}$. 
Then, the current information $(\bm{S}_j^{(b)}, \bm{U}_j^{(b)})$ and the prior information $(\bm{S}_j^{(b-1)}, \bm{U}_j^{(b-1)})$ and $(\bm{X}_j^{(b-1)}, \bm{Y}_j^{(b-1)})$ are combined to generate  the channel input $\bm{X}_j^{(b)}$. 

To reconstruct source messages, we employ a sliding-window decoder as depicted in Fig.~\ref{fig:DEC}. 
The decoder is designed to operate on two consecutive transmission blocks, but each time it only decodes the earlier source block. 
For $2\le b\le B+1$, suppose that the decoding window is now across the $(b-1)$\mbox{st} and the $b$th transmission blocks. 
Given that terminal~$j$ has successfully recovered $\bm{U}^{(b')}_{j'}$ and reconstructed $\bm{S}^{(b')}_{j'}$ for all $b'<b-1$, the decoder uses all available information in the $(b-1)$\mbox{st} and the $b$th blocks to recover $\bm{U}_{j'}^{(b-1)}$ and reconstructs $\bm{S}_{j'}^{(b-1)}$ as $\hat{\bm{S}}_{j'}^{(b-1)}$ via $G_j$. Then, the decoder moves to the $b$th and the $(b+1)$st blocks to reconstruct $\bm{S}_{j'}^{(b)}$.


With the above sketch, the left-hand-side and the right-hand-side of \eqref{eq:mainconds} can be interpreted as source compression rates and as transmission rates for reliable communication, respectively. 
Moreover, the appearance of $(\tilde{S}_j, \tilde{U}_j)$ (rather than $(S_j, U_j)$) on the left-hand-side of \eqref{eq:mainconds} is due to the sliding-window decoder. 
The tuple $(S_j, U_j, \tilde{S}_j, \tilde{U}_j, \tilde{W}_j, X_j, Y_j)$ on the right-hand-side of \eqref{eq:mainconds} also illuminates the fact that the decoder at terminal~$j$ uses all information within two blocks to decode $\tilde{U}_{j'}$. 
The detailed coding scheme and the formal proof is provided in Appendices~\ref{subsec:mainproof} and~\ref{subsec:claims}. 
In the next section, we simplify the expressions in \eqref{eq:mainconds} by imposing some encoding constraints. 
Examples illustrating the main theorem will be given in Section~\ref{sec:exdiss}.

\section{Simplified Configurations and Special Cases} \label{sec:speicalcases}
In this section, we consider two simplified forms of encoding to derive special cases from Theorem~\ref{thm:main}. 
Our objective is not only to obtain simpler achievability conditions but also to recover existing forward coding theorems for our problem setup. 
By-products of the derivation are reduced-complexity coding schemes in those special cases.  
As we will see later in Section~\ref{sec:JSCCthms}, the reduced-complexity schemes in the special cases are sometimes optimal in the sense that the associated achievable distortion region matches a certain outer bound; i.e., the scheme provides a complete JSCC theorem. 
In such a case, optimal performance can be achieved by a less complex coding scheme.
To ease our presentation, we will not refer to the probability distributions $P_{S_1, S_2}$ and $P_{Y_1, Y_2|X_1, X_2}$ in the following result statements as they are fixed and given by the problem setup. 
Also, we continue to focus on the rate-one case. 

\subsection{A Non-Adaptive JSCC Scheme}\label{subsec:twohybrid}
Our first simplification disables the superposition and adaptive coding components, i.e., we let $X_j=F_j(S_j, U_j, \tilde{S}_j, \tilde{U}_j, \tilde{W}_j)\triangleq f_j(\tilde{S}_j, \tilde{U}_j)$ and $\hat{\tilde{S}}_{j'}=G_j(\tilde{U}_{j'},  \allowbreak S_j, \allowbreak U_j, \allowbreak \tilde{S}_j, \allowbreak \tilde{U}_j, \allowbreak \tilde{W}_j, Y_j) \triangleq g_j(\tilde{U}_{j'}, \allowbreak \tilde{S}_j, \allowbreak \tilde{U}_j, \allowbreak Y_j)$ for some $f_j$ and $g_j$, $j=1, 2$. 
Set $P_{\tilde{S}_1, \tilde{S}_2}=P_{S_1, S_2}$, and set $P_{\tilde{U}_j|\tilde{S}_j}=P_{U_j|S_j}$ for a chosen $P_{U_j|S_j}$, $j=1, 2$, so that \eqref{eq:stationary1} and \eqref{eq:stationary2} holds.
We also set the pair $(\tilde{W}_1, \tilde{W}_2)$ to be {\it independent} of $(S_1, S_2, U_1, U_2, \tilde{S}_1, \tilde{S}_2, \tilde{U}_1, \tilde{U}_2, \allowbreak X_1, X_2, Y_1, Y_2)$ with joint probability distribution given by 
\begin{IEEEeqnarray}{l}
  P_{\tilde{W}_1, \tilde{W}_2}(\tilde{w}_1, \tilde{w}_2)=\sum_{a_1\in\mathcal{S}_1, a_2\in\mathcal{S}_2, b_1\in\mathcal{U}_1, b_2\in\mathcal{U}_2} P_{\tilde{S}_1, \tilde{S}_2}(a_1, a_2) P_{\tilde{U}_1|\tilde{S}_1}(b_1|a_1) P_{\tilde{U}_2|\tilde{S}_2}(b_2|a_2)\nonumber\\
  \quad\qquad\qquad\qquad\qquad\quad \mathbbm{1}\{\tilde{x}_1=f_1(a_1, b_1)\} \mathbbm{1}\{\tilde{x}_2=f_2(a_2, b_2)\} P_{Y_1, Y_2|X_1, X_2}(\tilde{y}_1, \tilde{y_2}|\tilde{x}_1, \tilde{x}_2).\IEEEeqnarraynumspace\label{eq:stathybrid}
\end{IEEEeqnarray}
With the above setting, one can directly verify that 
\begin{IEEEeqnarray}{l}
P_{Z}=P_{S_1, S_2}P_{U_1|S_1}P_{U_2|S_2}P_{\tilde{S}_1, \tilde{S}_2}P_{\tilde{U}_1|\tilde{S}_1}P_{\tilde{U}_2|\tilde{S}_2}P_{\tilde{W}_1, \tilde{W}_2}P_{X_1|\tilde{S}_1, \tilde{U}_1}P_{X_2|\tilde{S}_2, \tilde{U}_2}\allowbreak P_{Y_1, Y_2|X_1, X_2}\label{eq:cor1station}
\end{IEEEeqnarray} 
is a stationary distribution, i.e., $P_Z=Q_ZP_Z$. 
Given such $P_Z$, suppose that the chosen $g_j$ attains distortion level $D_j$, $j=1, 2$, so that the configuration $(P_{U_1|S_1}, P_{U_2|S_2}, P_{\tilde{S}_1, \tilde{S}_2, \tilde{U}_1, \tilde{U}_2}, P_{\tilde{W}_1, \tilde{W}_2}, f_1, f_2, \allowbreak g_1, g_2)$ is in $\Pi_Z(D_1, D_2)$. 
For simplicity, we define the set $\Pi'_Z(D_1, D_2)\subset\Pi_Z(D_1, D_2)$ as the one that contains all such special configurations.  
Using $\Pi'_Z(D_1, D_2)$, Theorem~\ref{thm:main} reduces to the following corollary. 



\begin{corollary}[Non-Adaptive Hybrid Coding]
  A distortion pair $(D_1, D_2)$ is achievable for the rate-one lossy transmission of correlated sources over a DM-TWC if there exists a configuration in $\Pi'_{Z}(D_1, D_2)$ such that
\label{cor:twchybrid}
\begin{subequations}
\label{eq:hybrid}
\begin{IEEEeqnarray}{rCl}
I(\tilde{S}_1; \tilde{U}_1|\tilde{S}_2, \tilde{U}_2) < I(\tilde{U}_1; Y_2|\tilde{S}_2, \tilde{U}_2),\label{eq:hybrida}\\
I(\tilde{S}_2; \tilde{U}_2|\tilde{S}_1, \tilde{U}_1) < I(\tilde{U}_2; Y_1|\tilde{S}_1, \tilde{U}_1).\label{eq:hybridb}
\end{IEEEeqnarray}
\end{subequations}
\end{corollary}
\begin{IEEEproof}
Since $\tilde{U}_{j'}$ is independent of $(S_j, U_j)$ and by definition $\tilde{W}_j$ is independent of $(\tilde{S}_{j'}, \allowbreak S_j, U_j, \tilde{S}_j, \tilde{U}_j, X_j, \allowbreak Y_j)$ for $j=1, 2$, we can remove $(S_j, U_j, \tilde{W}_j)$ from \eqref{eq:mainconds} without changing the values on the right-hand-side of \eqref{eq:mainconds}, e.g., \[I(\tilde{U}_1; S_2, U_2, \tilde{S}_2, \tilde{U}_2, \tilde{W}_2, X_2, Y_2)=I(\tilde{U}_1; \tilde{S}_2, \tilde{U}_2, X_2, Y_2)+\underbrace{I(\tilde{U}_1; S_2, U_2, \tilde{W}_2|\tilde{S}_2, \tilde{U}_2, X_2, Y_2)}_{=0}.\]
For \eqref{eq:mainconda}, we then have that 
\begin{IEEEeqnarray}{lrCl}
   & I(\tilde{S}_1; \tilde{U}_1) &< & I(\tilde{U}_1; \tilde{S}_2, \tilde{U}_2, X_2, Y_2)\nonumber\\
   \Leftrightarrow\ \   & H(\tilde{U}_1)-H(\tilde{U}_1|\tilde{S}_1) & < & I(\tilde{U}_1; \tilde{S}_2, \tilde{U}_2) + I(\tilde{U}_1; X_2, Y_2|\tilde{S}_2, \tilde{U}_2)\nonumber\\ 
   \Leftrightarrow\ \   & H(\tilde{U}_1)-H(\tilde{U}_1|\tilde{S}_1, \tilde{S}_2, \tilde{U}_2) & < & H(\tilde{U}_1)-H(\tilde{U}_1|\tilde{S}_2, \tilde{U}_2) + I(\tilde{U}_1; X_2, Y_2|\tilde{S}_2, \tilde{U}_2)\nonumber\\
   \Leftrightarrow\ \   & H(\tilde{U}_1|\tilde{S}_2, \tilde{U}_2)-H(\tilde{U}_1|\tilde{S}_1, \tilde{S}_2, \tilde{U}_2) & < & \underbrace{I(\tilde{U}_1; X_2|\tilde{S}_2, \tilde{U}_2)}_{=0} + \underbrace{I(\tilde{U}_1; Y_2|X_2, \tilde{S}_2, \tilde{U}_2)}_{=I(\tilde{U}_1; Y_2|\tilde{S}_2, \tilde{U}_2)}\label{eq:corpf1}\\
   \Leftrightarrow\ \   & I(\tilde{S}_1; \tilde{U}_1|\tilde{S}_2, \tilde{U}_2) & < & I(\tilde{U}_1; Y_2|\tilde{S}_2, \tilde{U}_2),\nonumber
\end{IEEEeqnarray}
where the two equalities in \eqref{eq:corpf1} hold since $X_2=f_2(\tilde{S}_2, \tilde{U}_2)$. 
By symmetry, one can analogously deduce \eqref{eq:hybridb} from \eqref{eq:maincondb}. 
\end{IEEEproof}

We remark that Corollary~\ref{cor:twchybrid} further subsumes several special cases. 
In the following derivations, we will show that our chosen parameters form a configuration in $\Pi'_Z(D_1, D_2)$. 
As $P_{\tilde{W}_1, \tilde{W}_2}$ can be determined via~\eqref{eq:stathybrid} given other parameters, we will not specify $P_{\tilde{W}_1, \tilde{W}_2}$ for the sake of simplicity. 

\begin{itemize}
  \item[(i)] \textbf{Uncoded transmission scheme}: Strictly speaking, the achievability result of an uncoded scheme cannot be deduced from Corollary~\ref{cor:twchybrid} since the conditions in \eqref{eq:hybrid} have no impact on the scheme's performance. 
  Nevertheless, we still can view it as a special case since every uncoded scheme can be converted into a configuration in our setup, which implies that our coding scheme (used to prove Theorem~\ref{thm:main}) can emulate uncoded transmission and attains the same distortion levels. 
  Specifically, let $\mathcal{X}_j=\mathcal{S}_j$, $j=1, 2$. 
  Given encoding functions $\tilde{f}_j$ and decoding functions $\tilde{g}_{j}$ 
  of an uncoded scheme such that $\mathbb{E}[d_{j}(\tilde{S}_{j}, \hat{\tilde{S}}_{j})]\le D_{j}$, we set $X_j=f_j(\tilde{U}_{j}, \tilde{S}_{j})=\tilde{f}_j(\tilde{S}_{j})$ and $\hat{\tilde{S}}_{j}=g_{j'}(\tilde{U}_{j}, \allowbreak \tilde{S}_{j'}, \allowbreak \tilde{U}_{j'}, \allowbreak Y_{j'})=\tilde{g}_{j'}(\tilde{S}_{j'}, Y_{j'})$.
  Also, set $P_{\tilde{S}_1, \tilde{S}_1}=P_{S_1, S_2}$ and $U_j=\tilde{U}_j=\text{constant}$. 
  This setting determines $P_{U_j|S_j}$ and $P_{\tilde{U}_j|\tilde{S}_j}$ uniquely and satisfies \eqref{eq:stationary1} and \eqref{eq:stationary2}. 
  We further obtain $P_{\tilde{W}_1, \tilde{W}_2}$ via \eqref{eq:stathybrid}. 
  Clearly, the configuration $(P_{U_1|S_1}, \allowbreak P_{U_2|S_2}, \allowbreak P_{\tilde{S}_1, \tilde{S}_2, \tilde{U}_1, \tilde{U}_2}, P_{\tilde{W}_1, \tilde{W}_2}, \tilde{f}_1,\allowbreak \tilde{f}_2, \allowbreak \tilde{g}_1, \tilde{g}_2)$ belongs to $\Pi'_Z(D_1, D_2)$. 
  Thus, one can establish the achievability result of uncoded transmission in our setup by giving appropriate functions $\tilde{f}_j$ and $\tilde{g}_j$. 
  A more detailed performance analysis for this scheme can be found in \cite{jjw2017}.  
  
  \item[(ii)] \textbf{SSCC for the lossy transmission of independent sources}: 
  To satisfy \eqref{eq:stationary1}, we let $P_{S_1, S_2}=P_{\tilde{S}_1, \tilde{S}_2}=P_{S_1}P_{S_2}$. 
  Define two independent random variables $V_1\in\mathcal{X}_1$ and $V_2\in\mathcal{X}_2$, whose joint probability distribution $P_{V_1}P_{V_2}$ achieves the rate pair $(I(V_1; Y_2|V_2), I(V_2; Y_1|V_1))$ in Shannon's capacity inner bound.
  For $j=1, 2$, we let $\hat{S}_j$ denote the reconstruction variable in the standard RD function of $S_j$  in \eqref{eq:RD} and choose $P_{\hat{S}_j|S_j}$ that attains $R^{(j)}(D_j)$. 
  Also, we define $(V'_1, V'_2)\in\mathcal{X}_1\times\mathcal{X}_2$ with $P_{V'_1}P_{V'_2}=P_{V_1}P_{V_2}$ and define $\hat{S}'_j\in\hat{\mathcal{S}}_j$ as the reconstruction variable in the standard RD function of $\tilde{S}_j$ at distortion level $D_j$, i.e., we set $P_{\hat{S}'_j|\tilde{S}_j}=P_{\hat{S}_j|S_j}$. 
  For $j=1, 2$, let $U_j\triangleq(V_j, \hat{S}_j)$ and $\tilde{U}_j\triangleq(V'_j, \hat{S}'_j)$ and set $P_{U_j|S_j}=P_{V_j}P_{\hat{S}_j|S_j}$ and $P_{\tilde{U}_j|\tilde{S}_j}=P_{V'_j}P_{\hat{S}'_j|\tilde{S}_j}$. 
  Clearly, the necessary condition in \eqref{eq:stationary2} is satisfied. 
  Moreover, set \[X_j=f_j(\tilde{U}_{j}, \tilde{S}_{j})=f_j((V'_j, \hat{S}'_j), \tilde{S}_{j})=V'_{j}\] 
  and choose the decoding function $g_{j}$ as \[\hat{\tilde{S}}_{j'}=g_{j}(\tilde{U}_{j'}, \tilde{U}_{j}, \tilde{S}_{j}, \tilde{Y}_{j})=g_{j}((V'_{j'}, \hat{S}'_{j'}), (V'_{j}, \hat{S}'_{j}), \tilde{S}_{j}, \tilde{Y}_{j})=\hat{S}'_{j'},\]
  which yields $\mathbb{E}[d_{j}(\tilde{S}_{j}, \hat{\tilde{S}}_{j})]\le D_{j}$ for $j=1, 2$.
  The above construction ensures that the tuple 
  \[(\underbrace{P_{V_1}P_{\hat{S}_1|S_1}}_{=P_{U_1|S_1}}, \underbrace{P_{V_2}P_{\hat{S}_2|S_2}}_{=P_{U_2|S_2}}, \underbrace{P_{\tilde{S}_1}P_{\tilde{S}_2}P_{V'_1}P_{\hat{S}'_1|\tilde{S}_1}P_{V'_2}P_{\hat{S}'_2|\tilde{S}_2}}_{=P_{\tilde{S}_1, \tilde{S}_2, \tilde{U}_1, \tilde{U}_2}}, P_{\tilde{W}_1, \tilde{W}_2}, f_1, f_2, g_1, g_2)\]
  is a configuration in $\Pi'_Z(D_1, D_2)$. 
  Next, using the fact that $S_1$ and $S_2$ are independent, one can simplify the sufficient conditions in \eqref{eq:hybrid} as follows (the details are given in Appendix~\ref{subsec:scii}):
  \begin{subequations}
    \begin{IEEEeqnarray}{rCl}
      R^{(1)}(D_1)&<& I(X_1; Y_2|X_2)\nonumber\\
      R^{(2)}(D_2)&<& I(X_1; Y_1|X_2)\nonumber
    \end{IEEEeqnarray}
  \end{subequations}
  which is the achievability result for the SSCC scheme based on the standard lossy source coding and Shannon's random channel coding (without time-sharing). \smallskip
  
  \item[(iii)] \textbf{SSCC for the lossy transmission of correlated sources}: 
  For $j=1, 2$, we define pairs $(V_1, V_2)\in\mathcal{X}_1\times\mathcal{X}_2$ and $(V'_1, V'_2)\in\mathcal{X}_1\times\mathcal{X}_2$ in the same way as in the special case (ii); set the two pairs to have identical distributions, i.e., $P_{V_1}P_{V_2}=P_{V'_1}P_{V'_2}$. 
  Letting $T_j\in\mathcal{T}_j$ denote the auxiliary random variable in the WZ RD function of $S_j$ in \eqref{eq:WZRD}, we choose $P_{T_j|S_j}$ and the associated decoding function $h_{j'}(T_j, S_{j'})$ that achieves $R_{\text{WZ}}^{(j)}(D_j)$. 
  Similarly, we use $T'_j\in\mathcal{T}_j$ in the WZ RD function of $\tilde{S}_j$ and set $P_{T'_j|\tilde{S}_j}=P_{T_j|S_j}$.  
  Letting $U_j\triangleq (V_j, T_j)$ and $\tilde{U}_j\triangleq (V'_j, T'_j)$, 
  we set $P_{U_j|S_j}=P_{V_j}P_{T_j|S_j}$ and $P_{\tilde{U}_j|\tilde{S}_j}=P_{V'_j}P_{T'_j|\tilde{S}_j}$. Also, set $P_{\tilde{S}_1, \tilde{S}_2}=P_{S_1, S_2}$. Thus, \eqref{eq:stationary1} and~\eqref{eq:stationary2} are satisfied. 
  Moreover, we set the encoding and decoding functions as 
  \[X_j=f_j(\tilde{U}_{j}, \tilde{S}_{j})=f_j((V'_{j}, T'_j), \tilde{S}_{j})=V'_{j}\] 
  and 
  \[\hat{\tilde{S}}_j=g_{j'}(\tilde{U}_{j}, \tilde{U}_{j'}, \tilde{S}_{j'}, \tilde{Y}_{j'})= g_{j'}((V'_{j}, T'_j), (V'_{j'}, T'_{j'}), \tilde{S}_{j'}, \tilde{Y}_{j'})=h_{j'}(T'_j, \tilde{S}_{j'}),\] 
  such that the decoder satisfies $\mathbb{E}[d_{j}(\tilde{S}_{j}, \hat{\tilde{S}}_{j})]\le D_{j}$ for $j=1, 2$. 
  With the above specifications, we next apply \eqref{eq:stathybrid} to obtain $P_{\tilde{W}_1, \tilde{W}_2}$, yielding the following configuration in $\Pi'_Z(D_1, D_2)$: 
  \[(\underbrace{P_{V_1}P_{T_1|S_1}}_{=P_{U_1|S_1}}, \underbrace{P_{V_2}P_{T_2|S_2}}_{=P_{U_2|S_2}}, \underbrace{P_{\tilde{S}_1, \tilde{S}_2}P_{V'_1}P_{T'_1|\tilde{S}_1}P_{V'_2}P_{T'_2|\tilde{S}_2}}_{=P_{\tilde{S}_1, \tilde{S}_2, \tilde{U}_1, \tilde{U}_2}}, P_{\tilde{W}_1, \tilde{W}_1}, f_1, f_2, h_1, h_2).\]

  Furthermore, using the Markov chain relationship: $T'_{1}\markov \tilde{S}_{1}\markov \tilde{S}_2\markov T'_2$ and the memoryless property of the channel, one can easily deduce the following two inequalities from \eqref{eq:hybrid}: 
  \begin{subequations}
    \begin{IEEEeqnarray}{rCl}
      R_{\text{WZ}}^{(1)}(D_1)&<&I(X_1; Y_2|X_2)\nonumber\\
      R_{\text{WZ}}^{(2)}(D_2)&<&I(X_2; Y_1|X_1)\nonumber
    \end{IEEEeqnarray}
  \end{subequations}
  which is the achievability result for the SSCC scheme based on the WZ lossy source coding and Shannon's random channel coding (without time-sharing) \cite{jjw2017}. 
  As the derivation is very similar to the previous case (see Appendix~\ref{subsec:scii}), we omit the details. 
  \smallskip
  
  \item[(iv)] \textbf{Correlation-preserving coding scheme for (almost) lossless transmission of correlated sources \cite{gunduz2009}}: 
  Suppose that $\mathcal{S}_j=\hat{\mathcal{S}}_j$ and consider the Hamming distortion measure \cite[Sec. 3.6]{kim2011}. 
  We first set $P_{\tilde{S}_1, \tilde{S}_2}=P_{S_1, S_2}$ to meet the necessary condition in \eqref{eq:stationary1}. 
  Recall the definitions of $(V_1, V_2)$ and $(V'_1, V'_2)$ in the special case (ii) with $P_{V_1}P_{V_2}=P_{V'_1}P_{V'_2}$, which achieve the same rate pair $(I(V_1; Y_2|V_2), I(V_2; Y_1|V_1))$ in Shannon's capacity inner bound.  
  Moreover, we recall the variables $(\hat{S}_1, \hat{S}_2)$ and $(\hat{S}'_1, \hat{S}'_2)$ from the special case (ii), but here we choose $P_{\hat{S}_j|S_j}$ to achieve $R^{(j)}(0)$ in \eqref{eq:RD} and set $P_{\hat{S}'_j|\tilde{S}_j}=P_{\hat{S}_j|S_j}$ for $j=1, 2$.   
  Let $U_j\triangleq (V_j, \hat{S}_j)$ and $\tilde{U}_j\triangleq (V'_j, \hat{S}'_j)$, and set $P_{U_j|S_j}=P_{V_j}P_{\hat{S}_j|S_j}$ and $P_{\tilde{U}_j|\tilde{S}_j}=P_{V'_j}P_{\hat{S}'_j|\tilde{S}_j}$. The setting satisfies the condition in \eqref{eq:stationary2}.
  We next consider the following encoding and decoding functions: \[X_j=f_j(\tilde{U}_{j}, \tilde{S}_{j})= f_j((V'_{j}, \hat{S}'_j), \tilde{S}_{j})=V'_{j}\] and \[\hat{\tilde{S}}_{j'}=g_{j}(\tilde{U}_{j'}, \tilde{U}_{j}, \tilde{S}_{j}, Y_{j})=g_{j}((V'_{j'}, \hat{S}'_{j'}), (V'_{j}, \hat{S}'_{j}), \tilde{S}_{j}, \tilde{Y}_{j})=\hat{S}'_{j'}.\] 
  Using \eqref{eq:stathybrid} to obtain $P_{\tilde{W}_1, \tilde{W}_2}$, we ensure that the resulting  configuration belongs to $\Pi'_Z(0, 0)$. 
  Furthermore, one can easily show that the sufficient conditions in \eqref{eq:hybrid} become 
  \begin{subequations}
    \begin{IEEEeqnarray}{rCl}
      R^{(1)}(0)=H(\tilde{S}_1|\tilde{S}_2)&<&I(V'_1; Y_2|V'_2, \tilde{S}_2)=I(X_1; Y_2|X_2, \tilde{S}_2)\nonumber\\
      R^{(2)}(0)=H(\tilde{S}_2|\tilde{S}_1)&<&I(V'_2; Y_1|V'_1, \tilde{S}_1)=I(X_2; Y_1|X_1, \tilde{S}_1)\nonumber
    \end{IEEEeqnarray}
  \end{subequations}
  which recover the achievability conditions in \cite[Cor.~8.1]{gunduz2009} (the rate-one case without coded time-sharing).
  Note that the block error rate for reconstructing the source messages is asymptotically vanishing here since the above conditions imply that $\lim_{K\to\infty}\Pr\big(\mathcal{E}\big)=0$ (see Appendix~\ref{subsec:mainproof} for the definition of the error event $\mathcal{E}$) and hence $\lim_{K\to\infty}\Pr\big((\tilde{S}_j^K, \hat{\tilde{S}}_j^K)\in\mathcal{T}_{\epsilon}^{(K)}\big)=1$ for $j=1, 2$, where $\mathcal{T}_{\epsilon}^{(K)}$ denotes the jointly typical set with parameters $K$ and $\epsilon$ as defined in \cite{kim2011}. 
  This result implies that $\lim_{K\to\infty}\Pr\big(\{\tilde{S}_1^K\neq\hat{\tilde{S}}_1^K\}\cup\{\tilde{S}_2^K\neq\hat{\tilde{S}}_2^K\} \big)=0$.
\end{itemize}\smallskip

\begin{figure}[!t]
  \centering
  \includegraphics[draft=false, scale=0.55]{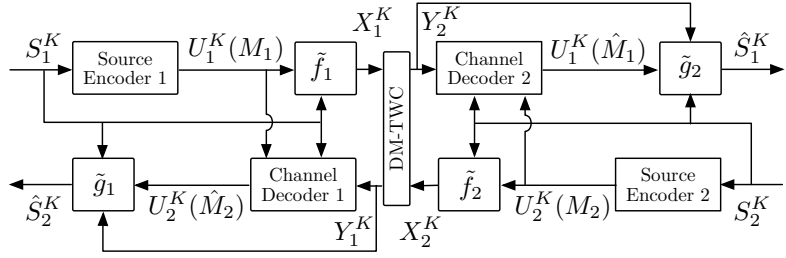}
  \caption{Rate-one non-adaptive hybrid coding scheme for the transmission of correlated sources over DM-TWCs.}
  \label{fig:Hybridblock}
 \end{figure}

In fact, since superposition coding is disabled in this simplified scheme, it is unnecessary to use the sliding window decoder. 
The decoding of each new source block can be done within the same transmission block. 
The block diagram of such coding system is depicted in Fig.~\ref{fig:Hybridblock} with the following system operations. 
The source messages $S_j^K$ are first mapped to a digital codeword $U_j^K(M_j)$ with index $M_j$. 
The channel inputs $X_j^K$ are then generated via the symbol-by-symbol map $\tilde{f}_j$, which combines the digital information $U_j^K(M_1)$ with the raw (or analog) information $S_j^K$.
Upon receiving $Y_j^K$, terminal~$j$ estimates the codeword index $M_{j'}$ based on all available information. 
Finally, the decoded codeword $U_{j'}(\hat{M}_{j'})$ and source message $S_j^K$ are passed together through the symbol-by-symbol map $\tilde{g}_j$ to produce $\hat{S}_{j'}^K$. 
The performance of this specific coding system is analyzed in \cite{jjw2019isit}. 
The sufficient conditions in the achievability result are identical to those in \eqref{eq:hybrid} except that $(\tilde{S}_1, \tilde{S}_2, \tilde{U}_1, \tilde{U}_2)$ are replaced with $(S_1, S_2, U_1, U_2)$. 
We remark that one can also employ the unified coding results in \cite{lee2018} to obtain these conditions since the coded system in Fig.~\ref{fig:Hybridblock} involves block-wise operations without adaptation. 

\subsection{An SSCC Scheme with Adaptive Channel Coding}
In the second simplification, we disable superposition coding for the raw source messages; i.e., 
we let $X_j=F_j(S_j, \allowbreak U_j, \allowbreak \tilde{S}_j, \allowbreak \tilde{U}_j, \allowbreak \tilde{W}_j)\triangleq f_j(U_j, \tilde{U}_j, \allowbreak \tilde{W}_j)$ and $\hat{\tilde{S}}_{j'}=G_j(\tilde{U}_{j'}, \allowbreak S_j, \allowbreak U_j, \allowbreak\tilde{S}_j, \allowbreak\tilde{U}_j, \allowbreak\tilde{W}_j, \allowbreak Y_j)\triangleq g_{j}(\tilde{U}_{j'}, \tilde{S}_j)$ for some $f_j$ and $g_j$, $j=1, 2$. 
Set $P_{\tilde{S}_1, \tilde{S}_2}=P_{S_1, S_2}$ to satisfy \eqref{eq:stationary1}.
Let $V_j$, $\tilde{V}_j$, and $\tilde{W}_j$ be the auxiliary random variables used in Han's result \cite{han1984} and let $\gamma_j: \mathcal{V}_j\times\tilde{\mathcal{V}}_j\times \tilde{\mathcal{W}}_j\to\mathcal{X}_j$ denote terminal $j$'s encoding function. 
Here, we choose $P_{V_1, V_2, \tilde{V}_1, \tilde{V}_2, \tilde{W}_1, \tilde{W}_2}$ and $\gamma_j$ that achieves the rate pair $(I(\tilde{V}_1; X_2, Y_2, \tilde{V}_2, \tilde{W}_2), I(\tilde{V}_2; X_1, Y_1, \tilde{V}_1, \tilde{W}_1))$ in Han's channel coding inner bound.
Note that in Han's result, $P_{V_1, V_2, \tilde{V}_1, \tilde{V}_2, \tilde{W}_1, \tilde{W}_2}=P_{V_1}P_{V_2}P_{\tilde{V}_1}P_{\tilde{V}_2}P_{\tilde{W}_1, \tilde{W}_2|\tilde{V}_1, \tilde{V}_2}$ and $P_{\tilde{V}_j}=P_{V_j}$, $j=1, 2$. 

Moreover, recall in \eqref{eq:WZRD} the auxiliary random variable $T_j$ in the WZ-RD function for $S_j$, $j=1, 2$; we choose $P_{T_j|S_j}$ and the associated decoding function $h_{j'}$ that attains $R_{\text{WZ}}^{(j)}(D_j)$. 
We also define its counterpart $\tilde{T}_j$ for $\tilde{S}_j$ and set $P_{\tilde{T}_j|\tilde{S}_j}=P_{T_j|S_j}$ for $j=1, 2$. 
Let $U_j\triangleq (V_j, T_j)$ and $\tilde{U}_j\triangleq (\tilde{V}_j, \tilde{T}_j)$ and set $P_{U_j|S_j}=P_{V_j}P_{T_j|S_j}$ and $P_{\tilde{U}_j|\tilde{S}_j}=P_{\tilde{V}_j}P_{\tilde{T}_j|\tilde{S}_j}$, which satisfy \eqref{eq:stationary2}. 
Next, we consider the following encoding and decoding functions: 
$f_j(U_j, \tilde{U}_j, \allowbreak \tilde{W}_j)=\gamma_j(V_j, \tilde{V}_j, \allowbreak \tilde{W}_j)$ and $g_{j}(\tilde{U}_{j'}, \tilde{S}_j)=h_{j}(\tilde{T}_{j'}, \tilde{S}_j)$, which ensures that $\mathbb{E}[d_{j}(\tilde{S}_{j}, \hat{\tilde{S}}_{j})]\le D_{j}$ for $j=1, 2$. 
Under the above setting, the joint probability distribution of all involved random variables is then given by 
\begin{IEEEeqnarray}{l}
  P_Z=P_{S_1, S_2}\underbrace{P_{V_1}P_{T_1|S_1}}_{=P_{U_1|S_1}}\underbrace{P_{V_2}P_{T_2|S_2}}_{=P_{U_2|S_2}}P_{\tilde{S}_1, \tilde{S}_2}\underbrace{P_{\tilde{V}_1}P_{\tilde{T}_1|\tilde{S}_1}}_{=P_{\tilde{U}_1|\tilde{S}_1}}\underbrace{P_{\tilde{V}_2}P_{\tilde{T}_2|\tilde{S}_2}}_{=P_{\tilde{U}_2|\tilde{S}_2}}\underbrace{P_{\tilde{W}_1, \tilde{W}_2|\tilde{V}_1, \tilde{V}_2}}_{=P_{\tilde{W}_1, \tilde{W}_2|\tilde{S}_1, \tilde{S}_2, \tilde{U}_1, \tilde{U}_2}}\nonumber\\
  \qquad\qquad\qquad\qquad \cdot P_{X_1|V_1, \tilde{V}_1, \tilde{W}_1}P_{X_2|V_2, \tilde{V}_2, \tilde{W}_2}P_{Y_1, Y_2|X_1, X_2},\label{eq:cor2station}
\end{IEEEeqnarray}
where $P_{\tilde{W}_1, \tilde{W}_2|\tilde{V}_1, \tilde{V}_2}$ is specified by Han's result \cite{han1984} and $P_{X_j|V_j, \tilde{V}_j, \tilde{W}_j}$ is determined by $\gamma_j$, $j= 1, 2$.    
It can be shown (by definition) that $P_Z=P_ZQ_Z$, thus implying that 
\[(P_{V_1}P_{T_1|S_1}, P_{V_2}P_{T_2|S_2}, P_{\tilde{S}_1, \tilde{S}_2}P_{\tilde{V}_1}P_{\tilde{T}_1|\tilde{S}_1}P_{\tilde{V}_2}P_{\tilde{T}_2|\tilde{S}_2}, P_{\tilde{W}_1, \tilde{W}_2|\tilde{V}_1, \tilde{V}_2}, \gamma_1, \gamma_2, h_1, h_2)\in \Pi_Z(D_1, D_2).\] 
Letting $\Pi{''}_Z(D_1, D_2)\subseteq \Pi_Z(D_1, D_2)$ denote the set of all such special configurations, we obtain the following corollary from Theorem~\ref{thm:main}.

\begin{corollary}[SSCC with WZ Source Coding and Han's Adaptive Channel Coding]
  \label{cor:ssccwzhan}
  A distortion pair $(D_1, D_2)$ is achievable for the rate-one lossy transmission of correlated sources over a DM-TWC if there exists a configuration in $\Pi^{''}_{Z}(D_1, D_2)$ such that
  \begin{subequations}
    \label{eq:ssccwzhan}
    \begin{IEEEeqnarray}{rCl}
      R^{(1)}_{\text{WZ}}(D_1)&< & I(\tilde{V}_1; X_2, Y_2, \tilde{V}_2, \tilde{W}_2),\label{eq:ssccwzhana}\\
      R^{(2)}_{\text{WZ}}(D_2)&< & I(\tilde{V}_2; X_1, Y_1, \tilde{V}_1, \tilde{W}_1).\label{eq:ssccwzhanb}
    \end{IEEEeqnarray}
    \end{subequations}
\end{corollary}
\begin{IEEEproof}
  For any configuration in $\Pi''_Z(D_1, D_2)$, the associated stationary distribution $P_{Z}$ can be factorized into the product form in \eqref{eq:cor2station}. 
  In addition to the independence between $(S_1, S_2, U_1, U_2)$ and $(\tilde{S}_1, \tilde{S}_2, \tilde{U}_1, \tilde{U}_2, \tilde{W}_1, \tilde{W}_2)$, the quadruple $(\tilde{S}_1, \tilde{S}_2, \tilde{T}_1, \tilde{T}_2)$ is independent of $(\tilde{V}_1, \tilde{V}_2)$. 
  These facts imply the independence between $\tilde{V}_j$ and $(S_{j'}, V_{j'}, \tilde{S}_{j'}, \tilde{T}_{j'})$. 
  Moreover, we have the following Markov chain relationships: $T_1\markov S_1 \markov S_2\markov T_2$, $\tilde{T}_1\markov \tilde{S}_1 \markov \tilde{S}_2\markov \tilde{T}_2$, and $\tilde{T}_j\markov (\tilde{V}_j, S_{j'}, U_{j'}, \tilde{S}_{j'}, \tilde{T}_{j'})\markov (\tilde{V}_{j'}, \tilde{W}_{j'}, X_{j'}, Y_{j'})$, $j=1, 2$.   
  We now show that \eqref{eq:mainconda} reduces to \eqref{eq:ssccwzhana}:  
  \begin{IEEEeqnarray}{lCrCl}
    & & I(\tilde{S}_1; \tilde{U}_1)&<& I(\tilde{U}_1; S_2, U_2, \tilde{S}_2, \tilde{U}_2, \tilde{W}_2, X_2, Y_2)\nonumber\\
    & \Leftrightarrow & I(\tilde{S}_1; \tilde{T}_1) + \underbrace{I(\tilde{S}_1; \tilde{V}_1|\tilde{T}_1)}_{=0} & < & \underbrace{I(\tilde{U}_1; S_2, U_2)}_{=0}+I(\tilde{U}_1; \tilde{S}_2, \tilde{V}_2, \tilde{T}_2, \tilde{W}_2, X_2, Y_2|S_2, U_2)\nonumber\IEEEeqnarraynumspace\\
    & \Leftrightarrow & I(\tilde{S}_1; \tilde{T}_1) - I(\tilde{U}_1; \tilde{S}_2, \tilde{T}_2|S_2, U_2) & < & I(\tilde{U}_1; \tilde{V}_2, \tilde{W}_2, X_2, Y_2|S_2, U_2, \tilde{S}_2, \tilde{T}_2)\nonumber\IEEEeqnarraynumspace\\
    & \Leftrightarrow & I(\tilde{S}_1; \tilde{T}_1) - I(\tilde{V}_1, \tilde{T}_1; \tilde{S}_2, \tilde{T}_2)& < & I(\tilde{V}_1, \tilde{T}_1; \tilde{V}_2, \tilde{W}_2, X_2, Y_2|S_2, U_2, \tilde{S}_2, \tilde{T}_2)\label{eq:x1}\\
    & \Leftrightarrow & I(\tilde{S}_1; \tilde{T}_1|\tilde{S}_2)& < & I(\tilde{V}_1; \tilde{V}_2, \tilde{W}_2, X_2, Y_2)\label{eq:x2}
  \end{IEEEeqnarray}
  where \eqref{eq:x1} holds since $I(\tilde{U}_1; \tilde{S}_2, \tilde{T}_2|S_2, U_2)=I(\tilde{U}_1; \tilde{S}_2, \tilde{T}_2)$ and $\tilde{U}_j=(\tilde{V}_j, \tilde{T}_j)$, and we have the equivalence in \eqref{eq:x2} since
  \begin{IEEEeqnarray}{l}
    I(\tilde{S}_1; \tilde{T}_1) - I(\tilde{V}_1, \tilde{T}_1; \tilde{S}_2, \tilde{T}_2) \nonumber\nonumber\\
    \ \ = I(\tilde{S}_1; \tilde{T}_1) - I(\tilde{T}_1; \tilde{S}_2, \tilde{T}_2)-\underbrace{I(\tilde{V}_1; \tilde{S}_2, \tilde{T}_2|\tilde{T}_1)}_{=0}\nonumber\\
    \ \ = I(\tilde{S}_1; \tilde{T}_1) - I(\tilde{T}_1; \tilde{S}_2, \tilde{T}_2) - I(\tilde{S}_1; \tilde{T}_1|\tilde{S}_2)+ I(\tilde{S}_1; \tilde{T}_1|\tilde{S}_2)\nonumber\\
    \ \ = H(\tilde{T}_1)-H(\tilde{T}_1|\tilde{S}_1)-H(\tilde{T}_1)+\underbrace{H(\tilde{T}_1|\tilde{S}_2, \tilde{T}_2)}_{=H(\tilde{T}_1|\tilde{S}_2)}-H(\tilde{T}_1|\tilde{S}_2)+\underbrace{H(\tilde{T}_1|\tilde{S}_1, \tilde{S}_2)}_{=H(\tilde{T}_1|\tilde{S}_1)}+ I(\tilde{S}_1; \tilde{T}_1|\tilde{S}_2)\nonumber\IEEEeqnarraynumspace\\
    \ \ = I(\tilde{S}_1; \tilde{T}_1|\tilde{S}_2),\nonumber 
  \end{IEEEeqnarray}
  and 
  \begin{IEEEeqnarray}{l}
    I(\tilde{V}_1, \tilde{T}_1; \tilde{V}_2, \tilde{W}_2, X_2, Y_2|S_2, U_2, \tilde{S}_2, \tilde{T}_2)\nonumber\\
    \ \ = I(\tilde{V}_1; \tilde{V}_2, \tilde{W}_2, X_2, Y_2|S_2, U_2, \tilde{S}_2, \tilde{T}_2) + \underbrace{I(\tilde{T}_1; \tilde{V}_2, \tilde{W}_2, X_2, Y_2|S_2, U_2, \tilde{S}_2, \tilde{T}_2, \tilde{V}_1)}_{=0}\nonumber\\
    \ \ = H(\tilde{V}_1|S_2, U_2, \tilde{S}_2, \tilde{T}_2) - H(\tilde{V}_1|S_2, U_2, \tilde{S}_2, \tilde{T}_2, \tilde{V}_2, \tilde{W}_2, X_2, Y_2)\nonumber\\
    \ \ = H(\tilde{V}_1) - H(\tilde{V}_1|\tilde{V}_2, \tilde{W}_2, X_2, Y_2)\label{eq:x3}\\
    \ \ = I(\tilde{V}_1; \tilde{V}_2, \tilde{W}_2, X_2, Y_2), \nonumber
  \end{IEEEeqnarray}
  where \eqref{eq:x3} holds since $\tilde{V}_1$ is independent of $(S_2, V_2, \tilde{S}_2, \tilde{T}_2)$ given $(\tilde{V}_2, \tilde{W}_2, X_2, Y_2)$. 
  By symmetry, one can also deduce \eqref{eq:ssccwzhanb} from \eqref{eq:maincondb}, thus completing the proof.
\end{IEEEproof}

We note that by working with super-symbols, we obtain a rate-$K/N$ extension of Corollary~\ref{cor:ssccwzhan}.
\begin{corollary}[General Rate SSCC with WZ Source Coding and Han's Adaptive Channel Coding]
\label{cor:sscc2}
A distortion pair $(D_1, D_2)$ is achievable for the rate-$K/N$ lossy transmission of correlated sources over a DM-TWC if
  \begin{subequations}
    \label{eq:sscc2}
    \begin{IEEEeqnarray}{rCl}
      K\cdot R^{(1)}_{\text{WZ}}(D_1)&< & N\cdot I(\tilde{V}_1; X_2, Y_2, \tilde{V}_2, \tilde{W}_2),\label{eq:sscc2a}\\
      K\cdot R^{(2)}_{\text{WZ}}(D_2)&< & N\cdot I(\tilde{V}_2; X_1, Y_1, \tilde{V}_1, \tilde{W}_1),\label{eq:sscc2b}
    \end{IEEEeqnarray}
  \end{subequations}
  for some joint probability distribution $P_{\tilde{V}_1, \tilde{V}_2, \tilde{W}_1, \tilde{W}_2, X_1, X_2}$ as defined in \cite[Section IV]{han1984}.
\end{corollary}

As Han's channel coding result subsumes Shannon's result, the following corollary is immediate, which is perhaps the simplest SSCC result for our problem setup. 

\begin{corollary}[General Rate SSCC with WZ Source Coding and Non-Adaptive Channel Coding]
  \label{cor:sscc1}
  A distortion pair $(D_1, D_2)$ is achievable for the rate-$K/N$ lossy transmission of correlated sources over a DM-TWC if
  \begin{subequations}
    \begin{IEEEeqnarray}{rCl}
      K\cdot R_{\text{WZ}}^{(1)}(D_1)< N\cdot I(X_1; Y_2|X_2),\\
      K\cdot R_{\text{WZ}}^{(2)}(D_2)< N\cdot I(X_2; Y_1|X_1),
    \end{IEEEeqnarray}
    \label{eq:sscc1}
  \end{subequations}
  for some $P_{X_1}P_{X_2}$.
\end{corollary}


We remark that since our general JSCC scheme (in the proof of Theorem~\ref{thm:main}) does not consider time-sharing for the sake of simplicity, the channel coding rate pairs obtained by the convex closure operation in Han's and Shannon's inner bound (see Section~\ref{subsec:CBs}) are excluded in Corollary~\ref{cor:sscc2} and Corollary~\ref{cor:sscc1}, respectively.
However, one can clearly incorporate time-sharing in our coding scheme and Theorem~\ref{thm:main}. 
After such convexification operation, one can include any achievable rate pair in Han's (resp., Shannon's) capacity inner bound region on the right-hand-side of \eqref{eq:sscc2} (resp., \eqref{eq:sscc1}). 
Furthermore, despite the fact that Corollary~\ref{cor:sscc2} strictly subsumes Corollary~\ref{cor:sscc1}, the associated achievable distortion regions are identical when DM-TWCs are symmetric \cite{jjw2019}; i.e., when Shannon's inner bound is tight.
In such situation, the simpler coding scheme of Corollary~\ref{cor:sscc1} is preferred.


\section{Converse Results and Complete JSCC Theorems}\label{sec:conandjssc}
The last two sections were devoted to the construction of achievable coding schemes. 
In this section, we derive two outer bounds to the achievable distortion region. 
Our objective is not only to identify unattainable distortion pairs but also to establish complete JSCC theorems.  

\subsection{Two Outer Bounds}
Lemmas~\ref{lma:OB1} and~\ref{lma:OB2} provide two outer bounds. 
Lemma~\ref{lma:OB2} is obtained via a genie-aided argument where the encoder at terminal~$j$ can access the decoder side-information $S_{j'}^K$ at terminal~$j'$. 
The proofs are standard and hence omitted. 
Details are given in \cite{jjw2017} and \cite{jjw2019isit}, respectively. 

\begin{lemma}\label{lma:OB1}
  If a rate-$K/N$ JSCC scheme achieves the distortion levels $D_1$ and $D_2$ for the lossy transmission of correlated sources over a DM-TWC, then 
  \begin{subequations}
  \begin{IEEEeqnarray}{rCl}
    K\cdot R^{(1)}(D_1)\le  K\cdot I(S_1; S_2) + N\cdot I(X_1; Y_2|X_2),\label{eq:OB1a}\\ 
    K\cdot R^{(2)}(D_2)\le  K\cdot I(S_1; S_2) + N\cdot I(X_2; Y_1|X_1),\label{eq:OB1b}
  \end{IEEEeqnarray}
  \end{subequations}
  for some $P_{X_1, X_2}$. 
  \label{eq:OB1}
\end{lemma}

\begin{lemma}[Genie-Aided Outer Bound]\label{lma:OB2}
  If a rate-$K/N$ JSCC scheme achieves the distortion levels $D_1$ and $D_2$ for the lossy transmission of correlated sources over a DM-TWC, then we have
  \begin{subequations}
  \begin{IEEEeqnarray}{rCl}
  K\cdot R_{S_1|S_2}(D_1)\le  N\cdot I(X_1; Y_2|X_2),\label{eq:OB2a}\\ 
  K\cdot R_{S_2|S_1}(D_2)\le  N\cdot I(X_2; Y_1|X_1),\label{eq:OB2b}
  \end{IEEEeqnarray}
  \end{subequations}
  for some $P_{X_1, X_2}$. 
  \label{eq:OB2}
\end{lemma}

Lemmas~\ref{lma:OB1} and~\ref{lma:OB2} generally give different outer bounds; however, the regions are identical for independent sources  $S_1$ and $S_2$ since in this case $I(S_1; S_2)=0$ and $R^{(j)}(D_j)=R_{S_j|S_{j'}}(D_j)$. 
The conditions in \eqref{eq:OB1} and \eqref{eq:OB2} are also equivalent for arbitrarily correlated sources for the specific distortion requirement $(D_1, D_2)=(0, 0)$ since $R_{S_j|S_{j'}}(0)=R^{(j)}(0)-I(S_1; S_2)=H(S_j|S_{j'})$. 


\subsection{Complete JSCC Theorems}\label{sec:JSCCthms}
Matching the achievability results in Section~\ref{sec:speicalcases} with the converse results in Lemmas~\ref{lma:OB1} and~\ref{lma:OB2}, we obtain three complete JSCC theorems (Theorems~\ref{thm:JSCC1}-\ref{thm:JSCC3}). 
We also establish a complete theorem (Theorem~\ref{thm:JSCC4}) for correlated source pairs that have common parts. 
In the results below, a ``symmetric DM-TWC'' is a DM-TWC that possesses the symmetry properties defined in \cite{jjw2019}. 
With these properties, Shannon's inner bound in \eqref{eq:shannonIB} is tight and hence the capacity region is achieved via independent inputs. 
Moreover, taking the convex closure in \eqref{eq:shannonIB} is not needed.

\begin{theorem}[Lossy Transmission of Indenpendent Sources]
  For the rate-$K/N$ lossy transmission of independent sources over a symmetric DM-TWC, a distortion pair $(D_1, D_2)$ is achievable if and only if 
  \vspace{-0.3cm}
  \begin{IEEEeqnarray}{rCl}
  K\cdot R^{(1)}(D_1)\le N\cdot I(X_1; Y_2|X_2),\nonumber\\
  K\cdot R^{(2)}(D_2)\le N\cdot I(X_2; Y_1|X_1),\nonumber
  \end{IEEEeqnarray}
  for some $P_{X_1}P_{X_2}$. 
  \label{thm:JSCC1}
\end{theorem}
\begin{IEEEproof}
This result is due to the special case (ii) of Corollary~\ref{cor:twchybrid} and Lemma~\ref{lma:OB1}, together with the facts that $R_{\text{WZ}}^{(j)}(D_j)=R^{(j)}(D_j)$ and $I(S_1; S_2)=0$ for independent sources pair. 
\end{IEEEproof}

\begin{theorem}[Almost Lossless Transmission of Correlated Sources]
  For the rate-$K/N$ transmission of correlated sources over a symmetric DM-TWC, the almost lossless transmission is achievable if and only if
  \vspace{-0.3cm}  
  \begin{IEEEeqnarray}{rCl}
  K\cdot H(S_1|S_2)\le N\cdot I(X_1; Y_2|X_2),\nonumber\\
  K\cdot H(S_2|S_1)\le N\cdot I(X_2; Y_1|X_1),\nonumber
  \end{IEEEeqnarray}
  for some $P_{X_1}P_{X_2}$. 
  \label{thm:JSCC2}
\end{theorem}
\begin{IEEEproof}
  In Lemma~\ref{lma:OB1}, we have that $K\cdot R^{(j)}(0)-K\cdot I(S_1; S_2) = K\cdot H(S_j|S_{j'})$. Combining this result with the special case (iv) of Corollary~\ref{cor:twchybrid} then completes the proof.   
\end{IEEEproof}

\begin{theorem}[Lossy Transmission of Correlated Sources with Equal WZ and Condtional RD Functions]  
  \label{thm:JSCC3}
  For the rate-$K/N$ lossy transmission of correlated sources whose WZ-RD functions equal to their conditional RD functions over a symmetric DM-TWC, a distortion pair $(D_1, D_2)$ is achievable if and only if  
  \vspace{-0.3cm}
  \begin{IEEEeqnarray}{rCl}
  K\cdot R_{S_1|S_2}(D_1)\le N\cdot I(X_1; Y_2|X_2),\nonumber\\
  K\cdot R_{S_2|S_1}(D_2)\le N\cdot I(X_2; Y_1|X_1),\nonumber
  \end{IEEEeqnarray}
  for some $P_{X_1}P_{X_2}$. 

\end{theorem}
\begin{IEEEproof}
  The result follows from the special case (iii) of Corollary~\ref{cor:twchybrid} and Lemma~\ref{lma:OB2}. 
\end{IEEEproof}

\begin{theorem}[Lossy Transmission of Correlated Sources with a Common Part]
  \label{thm:JSCC4}
  Assume that correlated sources $S_1$ and $S_2$ have a common part $S_0$ in the sense of G\'acs-K\"orner-Witsenhausen and the triplet $(S_0, S_1, S_2)$ forms a Markov chain $S_1\markov S_0\markov S_2$. For the rate-$K/N$ lossy transmission of such correlated sources over a symmetric DM-TWC, a distortion pair $(D_1, D_2)$ is achievable if and only if  
  \begin{subequations}
  \begin{IEEEeqnarray}{rCl}
  K\cdot R_{S_1|S_0}(D_1)\le N\cdot I(X_1; Y_2|X_2),\label{eq:OB3a}\\
  K\cdot R_{S_2|S_0}(D_2)\le N\cdot I(X_2; Y_1|X_1),\label{eq:OB3b}
  \end{IEEEeqnarray}
  \label{eq:OB3}
\end{subequations}
  for some $P_{X_1}P_{X_2}$. 
\end{theorem}
\begin{IEEEproof}
  We construct a two-way coding scheme using two one-way SSCC schemes, one for each direction of the bi-directional transmission. 
  Specifically, we employ the source coding scheme that achieves the distortion level $D_j$ of the conditional RD function $R^{(j)}_{S_j|S_0}(D_j)$ given in \eqref{eq:condRD}, $j=1, 2$, followed by Shannon's one-way channel coding for data protection.
  The sufficient conditions for achieving the distortion pair $(D_1, D_2)$ as shown in \eqref{eq:OB3} are thus immediate.
  Note that in this two-way coding scheme, we do not employ time-sharing and the channel inputs $X_1$ and $X_2$ are independent.

  The proof of the converse part is presented in Appendix~\ref{sub:jscc4converse}. 
  Although the inputs $X_1$ and $X_2$ are arbitrarily correlated in the outer bound result, we can restrict to independent inputs without changing the outer bound region due to the channel symmetry property, i.e., the capacity region of the DM-TWC can be determined via independent channel inputs.   
  Combining this fact with the achievability result then completes the proof. 
\end{IEEEproof}

\section{Examples and Discussion}\label{sec:exdiss}
\begin{figure}[!h]
  \centering
  \includegraphics[draft=false, scale=0.75]{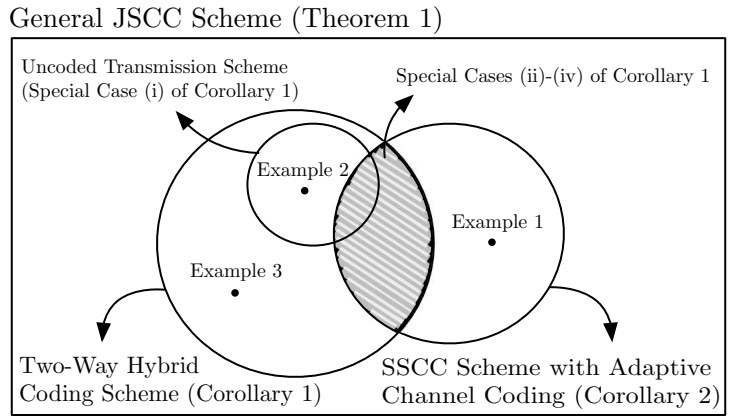}
  \caption{A general Venn diagram of the achievable distortion regions for the coding schemes presented  in Sections~\ref{sec:main} and~\ref{sec:speicalcases}, for a fixed source pair and channel. Moreover, Examples~1-3 in Section~\ref{subsec:examples} show that certain inclusion relationships can be strict.}
  \label{fig:distregion}
\end{figure}

In this section, we illustrate our achievability results and discuss possible extensions. 
The Venn diagram in Fig.~\ref{fig:distregion} summarizes the relationship of the achievable rate regions for the coding schemes in Sections~\ref{sec:main} and~\ref{sec:speicalcases}.   
We begin with three examples showing that some inclusion relationships can be strict, followed by illustrative examples for Theorems~\ref{thm:main},~\ref{thm:JSCC3}, and~\ref{thm:JSCC4}.    

\subsection{Examples}\label{subsec:examples}
Examples~1 and~2 below show that Theorem~\ref{thm:main} strictly generalizes Corollary~\ref{cor:twchybrid} and Corollary~\ref{cor:ssccwzhan}, respectively. 
Example~3 not only illustrates a special use of the two-way hybrid coding scheme but also reveals that Corollary~\ref{cor:twchybrid} strictly subsumes all of its special cases; see Section~\ref{subsec:twohybrid}. 
Example~4 shows how a simple instance of our adaptive JSCC helps source transmission. 
At the end of this section, we provide two examples (Examples 5-6) for Theorem~\ref{thm:JSCC3} and an example (Example 7) for Theorem~\ref{thm:JSCC4}. 
Note that except for the Gaussian case examined in Example~6, the Hamming distortion is considered in all examples. 
Let Ber$(p)$ denote a Bernoulli random variable with probability of success $p\in[0, 1]$, and let $H_{\text{b}}(\cdot)$ denotes the binary entropy function. 
We will also need the following specialized converse result in Examples~1 and~4, whose proof is similar to  Lemma~\ref{lma:OB1}.

\begin{proposition}
  \label{prop:converse1}
  Assume that the non-adaptive encoder $f_j: \mathcal{S}^{K}_j\to\mathcal{X}^{K}_j$ is used for $j=1, 2$.   
  If a distortion pair $(D_1, D_2)$ is achievable for the rate-one lossy transmission of independent sources over a DM-TWC, then
  \begin{IEEEeqnarray}{rCl}
  R^{(1)}(D_1)&\le & I(X_1; Y_2|X_2, Q),\nonumber\\
  R^{(2)}(D_2)&\le & I(X_2; Y_1|X_1, Q),\nonumber
  \end{IEEEeqnarray}
  for some $P_{Q}P_{X_1|Q}P_{X_2|Q}$.
\end{proposition}

Note that the pair $(I(X_1; Y_2|X_2, Q), I(X_2; Y_1|X_1, Q))$ under the distribution $P_{Q}P_{X_1|Q}P_{X_2|Q}$ in Proposition~\ref{prop:converse1} is an alternative expression for the achievable rate pair in Shannon's inner bound (see~\eqref{eq:shannonIB}).

\medskip
\begin{example}[\bf{Transmitting Independent Binary Non-Uniform Sources over Dueck's DM-TWC} \cite{dueck1979}] 
  Consider the independent sources $S_1=\text{Ber}(0.89)$ and $S_2=\text{Ber}(0.89)$ so that $H(S_1)=H(S_2)\approx 0.5$.
  We recall Dueck's DM-TWC \cite{dueck1979}, where $\bm{X}_j=(X_{j, 1}, X_{j, 2})$,\footnote{As Dueck's DM-TWC has $\mathcal{X}_j=\{0, 1\}^2$ and $\mathcal{Y}_j=\{0, 1\}^3$, we here use $(X_{j, 1}, X_{j, 2})\in\mathcal{X}_j$ to denote the two channel inputs of terminal $j$.} $\bm{Y}_j=(X_{1, 1}\cdot X_{2, 1}, N_j\oplus X_{j', 2}, N_{j'})$, the symbol $\oplus$ denotes the modulo-$2$ addition, and $N_1=\text{Ber}(0.5)$ and $N_2=\text{Ber}(0.5)$ are independent channel noise variables that are independent of all channel inputs and sources.
  Han \cite{han1984} showed that the channel coding rate pair $(R_{\text{c}, 1}, R_{\text{c}, 2})=(0.5, 0.5)$ is not achievable via Shannon's random coding scheme but can be achieved via his adaptive channel coding scheme. 
  Based on this fact and Proposition~\ref{prop:converse1}, we conclude that the hybrid coding scheme of Corollary~\ref{cor:twchybrid} cannot achieve the distortion pair $(D_1, D_2)=(0, 0)$ (since it uses non-adaptive encoders and violates the necessary conditions in Proposition~\ref{prop:converse1}).
  By contrast, Corollary~\ref{cor:ssccwzhan} shows that the distortion pair $(0, 0)$ is achievable via our general JSCC scheme as $R_{\text{WZ}, j}(0)=H(S_j)< R_{\text{c}, j}$ holds for $j=1, 2$. 
  Thus, Theorem~\ref{thm:main} strictly subsumes Corollary~\ref{cor:twchybrid}. 
\end{example}\medskip

\begin{example}[\bf{Transmitting Correlated Binary Sources over Binary-Multiplying DM-TWCs} \cite{shannon1961}]
  Consider the binary-multiplying TWC given by $Y_j=X_1\cdot X_{2}$ for $j=1, 2$.
  The capacity region of the channel is not known, but it is known that any symmetric achievable channel coding rate pair is component-wise upper bounded by $(0.646, 0.646)$ \cite{hekstra1989}. 
  Suppose that we want to exchange binary correlated sources with joint probability distribution $P_{S_1, S_2}(0, 0)=0$ and $P_{S_1, S_2}(s_1, s_2)=1/3$ for $(s_1, s_2)\neq (0, 0)$. 
  The WZ coding theorem indicates that the minimum source coding rate pair is $(H(S_1|S_2), H(S_2|S_1))=(0.667, 0.667)$ to achieve the distortion pair $(D_1, D_2)=(0, 0)$. 
  Clearly, this pair is not achievable by \emph{any} SSCC scheme, including the adaptive coding scheme of Corollary~\ref{cor:ssccwzhan}, because the source coding rate exceeds the largest possible transmission rate for reliable communication. 
  However, the uncoded scheme: $X_j=S_j$ for $j=1, 2$ can be easily shown to provide lossless transmission. 
  As Corollary~\ref{cor:ssccwzhan} and the uncoded scheme are special cases of our general JSCC method, Theorem~\ref{thm:main} strictly subsumes Corollary~\ref{cor:ssccwzhan}. 
\end{example} \medskip

\begin{example}[\bf{Transmitting Correlated Binary Sources over a Mixed-Type DM-TWC}]
  Suppose that all alphabets are binary. Let the source messages $S_1$ and $S_2$ have the joint probability distribution $P_{S_1, S_2}(1, 0)=0$  and $P_{S_1, S_2}(s_1, s_2)=1/3$ for $(s_1, s_2)\neq (1, 0)$. Consider the DM-TWC described by $Y_1=X_1\oplus X_2\oplus N_1$ and $Y_2=X_1\cdot X_2$, where $N_1=\text{Ber}(0.05)$ that is independent of $S_j$'s and $X_j$'s. 
  In other words, we have a (one-way) binary-multiplying channel in one direction and a binary additive channel with additive noise in another direction. 
  
  For this channel, none of the special cases of Corollary~\ref{cor:twchybrid} can achieve the distortion pair $(D_1, D_2)=(0, 0)$.
  More specifically, the SSCC schemes in the special cases cannot attain the distortion pair since $H(S_1|S_2)<I(X_1; Y_2|X_2)$ and $H(S_2|S_1)<I(X_2; Y_1|X_1)$ cannot hold simultaneously. 
  Moreover, using uncoded transmission in both directions yields the distortion pair $(D_1, D_2)=(0, 0.033)$. 
  However, we can use the two-way hybrid coding scheme in Corollary~\ref{cor:twchybrid} in the following way: use uncoded transmission from terminal~1 to~2 and use the concatenation of WZ source coding and Shannon's channel coding for the reverse direction. Then the distortion pair $(0, 0)$ is achievable. 
  This example shows that Corollary~\ref{cor:twchybrid} is a strictly generalization of its presented special cases.  
\end{example}\medskip

\begin{figure}[!t]
  \centering
  \includegraphics[draft=false, scale=0.59]{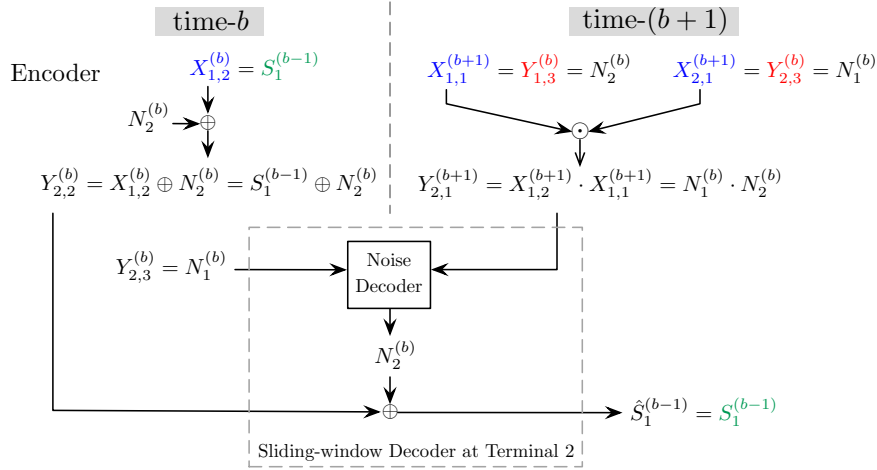}
  \caption{An illustration of adaptive encoding and sliding-window decoding in Example~4. At time-$b$, terminal~2 cannot perfectly decode $S_1^{(b-1)}$ from $Y_{2,2}^{(b)}$ due to the additive noise $N_2^{(b)}$. However, at time-$(b+1)$, the adaptive channel inputs $X_{1,1}^{(b+1)}$ and $X_{2,1}^{(b+1)}$ enable a perfect decoding for $N_2^{(b)}$ (based on $Y_{2,3}^{(b)}$, $Y_{2,1}^{(b+1)}$, and the noise correlation) at terminal~2, which can be used to eliminate the noise in $Y_{2,2}^{(b)}$ and achieve error-free transmission.}
  \label{fig:Ex4}
\end{figure}

\begin{example}[\bf{Transmitting Independent Binary Uniform Sources over Dueck's DM-TWC}]
  Consider the almost lossless transmission of the independent sources $S_1=\text{Ber}(0.5)$ and $S_2=\text{Ber}(0.5)$ through Dueck's DM-TWC (given in Example~1). Here, the binary noise variables $N_1$ and $N_2$ are assumed to be correlated with joint distribution given by $P_{N_1, N_2}(0, 0)=0$ and $P_{N_1, N_2}(n_1, n_2)=1/3$ for $(n_1, n_2)\neq (0, 0)$. 
  For this channel, the optimal symmetric rate pair in Proposition~\ref{prop:converse1} is obtained as $(I(X_1; Y_2|X_2), \allowbreak I(X_2; Y_1|X_1))=(0.9503, 0.9503)$.   
  Since the required source coding rate $R^{(j)}_{\text{WZ}}(0)=H(S_j)=1$ (at terminal~$j$) exceeds the outer bound in Proposition~\ref{prop:converse1}, the hybrid coding scheme in Corollary~\ref{cor:twchybrid} cannot achieve the distortion pair $(D_1, D_2)=(0, 0)$.
  
  By contrast, the following use of our general JSCC scheme provides rate-one lossless transmission. 
  Suppose that we exchange a length-$K$ of such source pair via $K+1$ channel uses. 
  Clearly, the transmission rate approaches one as $K$ goes to infinity. 
  For $j=1, 2$, we next set $(X^{(1)}_{j,1}, X^{(1)}_{j,2})=(1, S^{(1)}_{j})$, $(X^{(K+1)}_{j,1}, X^{(K+1)}_{j,2})=(Y^{(K)}_{j, 3}, 1)$, and $(X^{(b)}_{j,1}, X^{(b)}_{j,2})=(Y^{(n-1)}_{j, 3}, S^{(b)}_{j})$ for $b=2, 3, \dots, K$, where the superscripts represent time index. 
  Via such adaptive encoding, terminal~$j$ can exploit the correlation between $N_1$ and $N_2$ to perfectly decode $N^{(b-1)}_{j}$ from $Y^{(b)}_{j, 1}$ and $Y^{(b-1)}_{j, 3}$ and reconstruct $S^{(b-1)}_{j'}$ as $\hat{S}^{(b-1)}_{j'}=N^{(b-1)}_{j}\oplus Y^{(b-1)}_{j, 2}=S^{(b-1)}_{j'}$ for all $2\le b\le K+1$, thus achieving zero-error transmission. 
  For $2\le b\le K$, the above encoding and decoding procedure is depicted in Fig.~\ref{fig:Ex4}.  
  Note that whether or not the SSCC scheme in Corollary~\ref{cor:ssccwzhan} achieves the same performance remains unclear. 
\end{example}\medskip

\begin{example}[\bf{Transmitting Binary Correlated Sources with Z-channel Correlation over Binary Additive Noise DM-TWCs}]
  Suppose that all alphabets are binary. 
  Given $0\le \epsilon_1, \epsilon_2<0.5$, the binary additive noise DM-TWC is described by $Y_j=X_j\oplus X_{j'}\oplus N_j$, $j=1, 2$, where the channel noise variables $N_1=\text{Ber}(\epsilon_1)$ and $N_2=\text{Ber}(\epsilon_2)$ are independent of each other, of the source messages, and of the channel inputs. 
  The capacity region of the channel is given by \cite{song2016}: $\{(R_{\text{c}_1}, R_{\text{c}_2}):0\le R_{\text{c}_1}\le 1-H_{\text{b}}(\epsilon_2), 0\le R_{\text{c}_2}\le 1-H_{\text{b}}(\epsilon_1)\}$. 
  Consider the binary correlated source pair $(S_1, S_2)$ with Z-channel correlation \cite{nikos2014}; i.e., the transition matrices $[P_{S_2|S_1}(\cdot|\cdot)]$ and $[P_{S_1|S_2}(\cdot|\cdot)]$ between the sources $S_1$ and $S_2$ can be interpreted as a Z-channel and a reverse Z-channel, respectively. 
  Assume that the crossover probabilities of the Z-type channels are $\alpha_1$ and $\alpha_2$, respectively. 
  Let $P_{S_1}(1)=q_1$ and $P_{S_2}(1)=q_2$, where $q_2$ is a function of $q_1$ and $\alpha_1$ (note that one may also write $q_1$ as a function of $q_2$ and $\alpha_2$).  
  According to Theorem~\ref{thm:JSCC3}, the achievable distortion region for the rate-$K/N$ transmission consists of all pairs $(D_1, D_2)$ that satisfy the inequalities below:
  \begin{IEEEeqnarray}{rCl}
    K(1-q_1+q_1\alpha_1)\Bigg[H_{\text{b}}\bigg(\frac{q_1\alpha_1}{1-q_1+q_1\alpha_1}\bigg)-H_{\text{b}}\bigg(\frac{D_1}{1-q_1+q_1\alpha_1}\bigg)\Bigg]&\le& N(1-H_{\text{b}}(\epsilon_2)), \nonumber\\
    K(1-q_2+q_2\alpha_2)\Bigg[H_{\text{b}}\bigg(\frac{q_2\alpha_2}{1-q_2+q_2\alpha_2}\bigg)-H_{\text{b}}\bigg(\frac{D_2}{1-q_2+q_2\alpha_2}\bigg)\Bigg]& \le& N(1-H_{\text{b}}(\epsilon_1)). \nonumber
  \end{IEEEeqnarray}
\end{example}\bigskip

\begin{example}[\bf{Transmitting Correlated Gaussian Sources over DM-TWCs with Additive White Gaussian Noise (AWGN) DM-TWCs}]
  Consider the squared-error distortion measure. 
  The AWGN DM-TWC is described by $Y_j = X_j+X_{j'}+N_j$, $j=1, 2$, where $N_1$ and $N_2$ are independent zero mean Gaussian noises with variance $\sigma_1^2$ and $\sigma_2^2$, respectively, and are independent of the source messages and of the channel inputs. 
  The average power of channel inputs $X_j$ is set as $P_j$ for $j=1, 2$. 
  Moreover, the correlated sources $S_1$ and $S_2$ are considered to be zero-mean unit-variance jointly Gaussian random variables with correlation coefficient $\rho$ for some $0\le \rho\le 1$. 
  For this setting, Theorem~\ref{thm:JSCC3} yields the achievable distortion region $\{(D_1, D_2): D_j\ge (1-\rho^2)(1+\frac{P_j}{\sigma_{j'}})^{\frac{K}{N}}, j=1, 2\}$, for the rate-$K/N$ transmission. 
  The detailed derivation can be found in \cite[Lemma~4]{jjw2017}. 
\end{example}\medskip

\begin{example}[\bf{Transmitting Quaternary Correlated Sources over Binary Additive Noise DM-TWCs}]
Suppose that $\mathcal{S}_1=\mathcal{S}_2=\hat{\mathcal{S}}_1=\hat{\mathcal{S}}_2=\{A, B, C, D\}$ and $\mathcal{X}_1=\mathcal{X}_2=\mathcal{Y}_1=\mathcal{Y}_2=\{0, 1\}$. 
Consider the correlated source pair with joint probability distribution given by 
\begin{equation}
  P_{S_1, S_2}(s_1, s_2) = \left\{ \,
  \begin{IEEEeqnarraybox}[][c]{l?s}
  \IEEEstrut
  \frac{1}{8} & if $(s_1, s_2)\in\{A, B\}\times\{A, B\}\cup\{C, D\}\times\{C, D\}$, \\
  0 & otherwise.\nonumber
  \IEEEstrut
  \end{IEEEeqnarraybox}
  \right.
  \end{equation}
  For such sources, we observe a binary common part $S_0$; $S_0=0$ and $S_0=1$ are corresponding to $S_1, S_2 \in\{A, B\}$ and $S_1, S_2 \in\{C, D\}$, respectively. 
  Given this common part, we can decompose $S_j$ into $(S_0, S'_j)$, where $S'_j=\text{Ber}(0.5)$. 
  It is easy to show that $S_j$ and $(S_0, S'_j)$ have a one-to-one correspondence and the Markov chain relationship $S'_1\markov S_0\markov S'_2$ holds. 
  Moreover, the conditional RD function $P_{S'_j|S_0}(D_j)$ is given by $P_{S'_j|S_0}(D_j)=1-H_{\text{b}}(D_j)$ for $0\le D_j\le 0.5$. 

  Due to the above decomposition, the terminals only need to exchange $(S'_1, S'_2)$.  
  When transmitting the pair $(S'_1, S'_2)$ over the binary additive noise DM-TWCs (defined in Example 5) at rate-$K/N$, we can apply Theorem~\ref{thm:JSCC4} to characterize the achievable distortion region of the overall system, which is the convex hull of all distortion pairs $(D_1, D_2)$ satisfying
  \begin{IEEEeqnarray}{rCl}
      K(1-H_{\text{b}}(D_1))&\le & N(1-H_{\text{b}}(\epsilon_2)),\nonumber\\
      K(1-H_{\text{b}}(D_2))&\le & N(1-H_{\text{b}}(\epsilon_1)).\nonumber
  \end{IEEEeqnarray}
\end{example}


\subsection{Adaptive Coding with More Past Information}\label{subsec:disshist}
In our JSCC scheme (detailed in Appendix~\ref{subsec:mainproof}), we merely use the most recent channel inputs and outputs $(X^{(t-1)}_j, Y^{(t-1)}_j)$ to generate the current channel input $X^{(t)}_j$. 
Although ideally one would use the entire past channel input and output history for adaptive coding, the accumulated information in this case causes the Markov chain not only to have a time-varying transition kernel but also to drastically expand the state space. 
The idea to jointly optimize the terminals' transmission via a stationary Markov chain becomes infeasible. 
In the following, we sketch two coding strategies to deal with this problem. 
Each of the strategies can be directly integrated into our JSCC scheme, but the encoding/decoding complexity will be higher and the sufficient conditions will be significantly more complicated than the current ones. 

The first strategy is to generate $X^{(t)}_j$ as a function of the past $\mu$ channel inputs $(X^{(t-\mu)}_j, X^{(t-\mu+1)}_j, \allowbreak\dots, X^{(t-1)}_j )$ and outputs $(Y^{(t-\mu)}_j, Y^{(t-\mu+1)}_j, \dots, Y^{(t-1)}_j )$ for some $\mu >1$, which is similar to the memory-$\mu$ channel coding for DM-TWCs \cite[Section 4.4]{kramer1998}. 
This strategy increases the encoding and decoding complexity, but the state space complexity of the Markov chain is constant.  

The second strategy quantizes the past channel inputs and outputs at each terminal into a set with fixed size. 
The channel inputs can be then generated as a function of the quantized information in that set, rather than the entire past information. 
This strategy is similar to the Q-graph channel coding for single-output DM-TWCs \cite{sabag2018}, and it adds a minor encoding cost. 
However, as the quantized knowledge is not necessarily a sufficient statistic for optimal decoding, we still need to store all past information, which clearly increases system complexity.\footnote{One can apply sliding-window decoding to limit the amount of past information at each receiver.}

\subsection{Adaptive Coding with Incremental Side-Information}\label{subsec:disssource}
Our adaptive coding mainly coordinates the terminals' transmission on the shared channel as we did not attempt to apply Kaspi's interactive source coding idea \cite{kaspi1985} to make the best use of the sequentially received signals.  
Here, we give an SSCC scheme that encompasses both ideas. 

The exchange of correlated sources $S_1^K$ and $S_2^K$ is now accomplished in $L$ rounds for some $L\ge 1$, which comprises $N$ channel uses (note that $N$ is a function of $K$).
Specifically, for $1\le l\le L$, let $N_l$ denote the number of channel uses in the $l$th round of transmission, where $\sum_{l=1}^L N_l=N$. 
In each round, viewing the previously transmitted and decoded source codewords as side-information, each terminal applies binning for source coding, followed by Han's adaptive channel coding. 
Each terminal also decodes the other terminal's source codeword at the end of each transmission round. 
After $L$ rounds, each terminal reconstructs the other terminal's source messages from the side-information and its own source messages.  
Clearly, this simple SSCC scheme allows two-way simultaneous transmission and interactive source coding. 
We summarize the achievability result in Proposition~2 below (without proof). Here, $T_{j,l}$, $j=1, 2$ and $l=1, 2, \dots, L$, are auxiliary random variables.   

\begin{proposition}
  A distortion pair $(D_1, D_2)$ is achievable for the rate-$K/N$ lossy transmission of correlated sources over a DM-TWC if for all $1\le l\le L$, we have that
  \begin{subequations}
    \begin{IEEEeqnarray}{rCl}
      K\cdot I(S_1; T_{1,l}|S_2, T_1^{l-1}, T_2^{l-1})< N_l\cdot I(\tilde{V}_{1,l}; X_{2,l}, Y_{2,l}, \tilde{V}_{2,l}, \tilde{W}_{2,l}),\nonumber\\
      K\cdot I(S_2; T_{2,l}|S_1, T_1^{l-1}, T_2^{l-1})< N_l\cdot I(\tilde{V}_{2,l}; X_{1,l}, Y_{1,l}, \tilde{V}_{1,l}, \tilde{W}_{1,l}),\nonumber
    \end{IEEEeqnarray}
  \end{subequations}
   for some joint probability distributions $P_{\tilde{V}_{1,l}, \tilde{V}_{2,l}, \tilde{W}_{1,l}, \tilde{W}_{2,l}, X_{1,l}, X_{2,l}}$ as defined in \cite[Section IV]{han1984} and 
   \[
   P_{T_1^L, T_2^L|S_1, S_2}=\prod_{l=1}^L P_{T_{1, l}|S_1, T_1^{l-1}, T_2^{l-1}} P_{T_{2, l}|S_2, T_1^{l-1}, T_2^{l-1}} 
   \] 
   and two decoding functions $\hat{S}_{j'}=g_j(S_j, T_j^L, T^L_{j'})$ such that $\mathbb{E}[d_j(S_j, \hat{S}_j)]\le D_j$ for $j=1, 2$. 
\end{proposition}

Note that the above proposition reduces to Corollary~\ref{cor:sscc2} when $L=1$. In light of this, it is of interest to ask if there exists a general adaptive JSCC scheme that integrates both features and subsumes all of our presented achievability results.
We leave this question for future research. 

\section{Conclusions}\label{sec:conclusion}
We constructed an adaptive coding scheme to prove a forward JSCC theorem, which characterizes an achievable distortion region for two-way lossy simultaneous transmission.  
Our adaptive coding method demonstrates a way to coordinate the independent transmissions of the terminals; it also underscores the importance of preserving source correlation as illustrated via several examples. 
Moreover, our coding scheme subsumes several simple non-adaptive coding methods, providing a unified transmission framework that allows for diverse various system complexity and performance trade-offs.
Although the general form of our scheme is complex, in many cases its SSCC instances suffice to achieve the optimal performance. 
Future directions include adaptive coding based on the SSCC structure, symbol-wise adaptive coding (as opposed to block-wise adaptive coding), and practical joint source-channel code design for our problem setup. 
It is also of interest to refine the outer bounds and derive a complete characterization of the achievable RD region for two-way source-channel communication (in either single-letter or multi-letter expression). 

\begin{appendix}
\subsection{Proof of Theorem~\ref{thm:main}}\label{subsec:mainproof}
  For the sake of brevity, the complete proof is presented in this section using several auxiliary claims whose proofs are given in Appendix~\ref{subsec:claims}. 
  Let $\mathcal{T}_{\epsilon}^{(n)}$ denote the typical set of sequences with parameters $n\in\mathbb{Z}_{+}$ and $\epsilon>0$ as defined in \cite{kim2011}; the domain of $\mathcal{T}_{\epsilon}^{(n)}$ will clear from the context and hence omitted. 
  Here, we set $n=N=K$ as we consider the rate-one transmission. 
  For $j=1, 2$ and $b=1, 2, \cdots, B$, we define $2^{\mli{nR}^{(b)}_j}$ as the size of terminal~$j$'s codebook $\mathcal{C}^{(b)}_j$, which is used to encode the $b$-th block $\bm{S}^{(b)}_j$ of source messages.
  For an event $\mathcal{E}$, we let $\overline{\mathcal{E}}$ denote its complement. 
  \smallskip
  
  \noindent\underline{\emph{Codebook Generation}}: Given a configuration in $\Pi_Z(D_1, D_2)$, generate two length-$n$ sequences $(\tilde{\bm{S}}^{(1)}_1, \allowbreak\tilde{\bm{S}}^{(1)}_2, \allowbreak\tilde{\bm{U}}^{(1)}_1, \allowbreak\tilde{\bm{U}}^{(1)}_2, \allowbreak\tilde{\bm{W}}^{(1)}_1, \allowbreak\tilde{\bm{W}}^{(1)}_2)$ and $(\bm{S}^{(B+1)}_1, \allowbreak\bm{S}^{(B+1)}_2, \allowbreak\bm{U}^{(B+1)}_1, \allowbreak\bm{U}^{(B+1)}_2)$ to initialize and terminate the $(B+1)$-blocks encoding process with distributions 
  \begin{IEEEeqnarray}{l}
  \scalemath{0.95}{P_{\tilde{\bm{S}}^{(1)}_1, \tilde{\bm{S}}^{(1)}_2, \tilde{\bm{U}}^{(1)}_1, \tilde{\bm{U}}^{(1)}_2, \tilde{\bm{W}}^{(1)}_1, \tilde{\bm{W}}^{(1)}_2}(\tilde{\bm{s}}^{(1)}_1, \tilde{\bm{s}}^{(1)}_2, \tilde{\bm{u}}^{(1)}_1, \tilde{\bm{u}}^{(1)}_2, \tilde{\bm{w}}^{(1)}_1, \tilde{\bm{w}}^{(1)}_2)}\nonumber\\
  \qquad\qquad\qquad\qquad\qquad\ \scalemath{0.95}{=\prod_{i=1}^n P_{\tilde{S}_1, \tilde{S}_2, \tilde{U}_1, \tilde{U}_2, \tilde{W}_1, \tilde{W}_2}(\tilde{s}^{(1)}_{1, i}, \tilde{s}^{(1)}_{2, i}, \tilde{u}^{(1)}_{1, i}, \tilde{u}^{(1)}_{2, i}, \tilde{w}^{(1)}_{1, i}, \tilde{w}^{(1)}_{2, i})}\label{eq:cwgen1}
  \end{IEEEeqnarray}
  and
  \begin{IEEEeqnarray}{l}
  \scalemath{0.95}{P_{\bm{S}^{(B+1)}_1, \bm{S}^{(B+1)}_2, \bm{U}^{(B+1)}_1, \bm{U}^{(B+1)}_2}(\bm{s}^{(B+1)}_1, \bm{s}^{(B+1)}_2, \bm{u}^{(B+1)}_1, \bm{u}^{(B+1)}_2)}\nonumber\\
  \qquad\qquad\qquad\qquad\quad\scalemath{0.95}{=\prod_{i=1}^n P_{S_1, S_2, U_1, U_2}(s^{(B+1)}_{1, i}, s^{(B+1)}_{2, i}, u^{(B+1)}_{1, i}, u^{(B+1)}_{2, i})}. \label{eq:cwgen2}
  \end{IEEEeqnarray}
  Moreover, generate codebooks $\mathcal{C}^{(b)}_j\triangleq\{\bm{U}^{(b)}_j(m^{(b)}_j): m^{(b)}_j=1, 2, \dots, 2^{\mli{nR^{(b)}_j}}\}$ for $b=1, 2, \dots, B$ and $j=1, 2$, where $\bm{U}^{(b)}_j(m^{(b)}_j)$ is a length-$n$ sequence distributed according to $P_{\bm{U}_j}(\bm{u}^{(b)}_j(m^{(b)}_j))=\allowbreak\prod_{i=1}^n P_{U_j}(u^{(b)}_{j, i}(m^{(b)}_j))$ and $\bm{U}^{(b)}_j(m^{(b)}_j)$'s are independent of each other. 
  The initialization and termination sequences and all codebooks are revealed to both terminals.
  We note that due to the construction of the Markov chain $\{Z^{(t)}\}$, the codebook $\mathcal{C}^{(b)}_j$ is also used for $\tilde{\bm{U}}_j^{(b+1)}$.\smallskip
  
  \noindent\underline{\emph{Encoding}}: Let $\epsilon_1>\epsilon>0$. For $b=1, 2, \dots, B$ and $j=1, 2$, terminal~$j$ finds $m^{(b)}_j$ such that $(\bm{S}^{(b)}_j, \bm{U}(m^{(b)}_j))\in\mathcal{T}_{\epsilon_1}^{(n)}$. If there is more than one such index, the encoder chooses one of them at random. If there is no such index, it chooses an index at random from $\{1, 2, \dots, 2^{\mli{nR_j^{(b)}}}\}$. The transmitter then sends $\bm{X}^{(b)}_j$, where $X^{(b)}_{j, i}=F_j(S_{j, i}^{(b)}, U_{j, i}^{(b)}(m_j^{(b)}), \tilde{S}_{j,i}^{(b)}, \tilde{U}_{j, i}^{(b)}, \tilde{W}_{j, i}^{(b)})$ for $i=1, 2, \dots, n$, $\tilde{S}_{j,i}^{(b)}=S_{j,i}^{(b-1)}$, $\tilde{U}_{j, i}^{(b)}=U_{j, i}^{(b-1)}$, and $\tilde{W}_{j, i}^{(b)}=(X_{j, i}^{(b-1)}, Y_{j, i}^{(b-1)})$ for $b=2, 3, \dots, B$.
  For $b=B+1$, $\bm{X}^{(B+1)}$ is generated in the same way using the termination sequence.\smallskip
  
  \noindent\underline{\emph{Decoding}}: For $b{=}2, 3, \dots, B+1$ and $j, j'{=}1, 2$ with $j{\neq} j'$, terminal~$j$ finds an index $\hat{m}^{(b-1)}_{j'}$ such that 
  $(\bm{S}_j^{(b)}, \allowbreak\bm{U}_j^{(b)}, \allowbreak\tilde{\bm{S}}_j^{(b)}, \allowbreak\tilde{\bm{U}}_j^{(b)}, \allowbreak\tilde{\bm{U}}_{j'}^{(b)}(\hat{m}^{(b-1)}_{j'}), \allowbreak\tilde{\bm{W}}^{(b)}_j, \bm{X}^{(b)}_j, \allowbreak\bm{Y}^{(b)}_j)\in\mathcal{T}^{(n)}_{\epsilon},$    
  where $\tilde{\bm{U}}_{j'}^{(b)}(\hat{m}^{(b-1)}_{j'})\in\mathcal{C}_{j'}^{(b-1)}$. If there is more than one choice, the decoder chooses one of them at random. 
  If there is no such index, it chooses one at random from $\{1, 2, \allowbreak\dots, \allowbreak 2^{\mli{nR_{j'}^{(b)}}}\}$.
  The reconstruction for the source message $\bm{S}_{j'}^{(b-1)}$ is given by $\hat{S}^{(b-1)}_{j', i}=G_j(\tilde{U}_{j', i}^{(b)}(\hat{m}^{(b-1)}_{j'}), \allowbreak S^{(b)}_{j, i}, \allowbreak U^{(b)}_{j, i}, \allowbreak \tilde{S}_{j, i}^{(b)}, \allowbreak\tilde{U}_{j, i}^{(b)}, \allowbreak\tilde{W}^{(b)}_{j, i}, \allowbreak Y^{(b)}_{j, i})$ for $i=1, 2, \dots, n$.\smallskip
  
  \begin{figure*}[b]
    \hrulefill
    \begin{IEEEeqnarray}{rCl}
      \label{eq:errorevent}
      \IEEEyesnumber
      \IEEEyessubnumber*
      \scalemath{0.92}{\mathcal{E}_1^{(1)}}&\triangleq &\scalemath{0.92}{\{(\bm{S}_1^{(1)}, \bm{S}_2^{(1)}, \bm{U}_1^{(1)}(M_1^{(1)}), \bm{U}_2^{(1)}(M_2^{(1)}), \tilde{\bm{S}}_1^{(1)}, \tilde{\bm{S}}_2^{(1)},} 
      \nonumber\\
      & & \qquad\qquad\qquad\qquad\qquad\scalemath{0.92}{\tilde{\bm{U}}_1^{(1)}, \tilde{\bm{U}}_2^{(1)}, \tilde{\bm{W}}_1^{(1)}, \tilde{\bm{W}}_2^{(1)}, \bm{X}_1^{(b)}, \bm{X}_2^{(b)}, \bm{Y}_1^{(b)}, \bm{Y}_2^{(b)})\notin T_{\epsilon}^{(n)}\}}.\IEEEeqnarraynumspace\\
      \scalemath{0.92}{\mathcal{E}_1^{(B+1)}}&\triangleq &\scalemath{0.92}{\{(\bm{S}_1^{(B+1)}, \bm{S}_2^{(B+1)}, \bm{U}_1^{(B+1)}, \bm{U}_2^{(B+1)}, \tilde{\bm{S}}_1^{(B+1)}, \tilde{\bm{S}}_2^{(B+1)}, \tilde{\bm{U}}_1^{(B+1)}(\hat{M}_1^{(B)}), \tilde{\bm{U}}_2^{(B+1)}(M_2^{(B)}), } \nonumber\\
      & & \qquad\qquad\qquad\qquad\scalemath{0.92}{\tilde{\bm{W}}_1^{(B+1)}, \tilde{\bm{W}}_2^{(B+1)}, \bm{X}_1^{(B+1)}, \bm{X}_2^{(B+1)}, \bm{Y}_1^{(B+1)}, \bm{Y}_2^{(B+1)})\notin T_{\epsilon}^{(n)}\}}.\IEEEeqnarraynumspace\\
      \scalemath{0.92}{\mathcal{E}_1^{(b)}}&\triangleq &\scalemath{0.92}{\{(\bm{S}_1^{(b)}, \bm{S}_2^{(b)}, \bm{U}_1^{(b)}(M_1^{(b)}), \bm{U}_2^{(b)}(M_2^{(b)}), \tilde{\bm{S}}_1^{(b)}, \tilde{\bm{S}}_2^{(b)},\tilde{\bm{U}}_1^{(b)}(\hat{M}_1^{(b-1)}), \tilde{\bm{U}}_2^{(b)}(M_2^{(b-1)}),} \nonumber\\
      & & \qquad\qquad\qquad\quad\scalemath{0.92}{\tilde{\bm{W}}_1^{(b)}, \tilde{\bm{W}}_2^{(b)}, \bm{X}_1^{(b)}, \bm{X}_2^{(b)}, \bm{Y}_1^{(b)}, \bm{Y}_2^{(b)})\notin T_{\epsilon}^{(n)}\},\text{\ for\ } b=2, 3, \dots, B.}\IEEEeqnarraynumspace 
    \end{IEEEeqnarray}
  \end{figure*}

  \noindent\underline{\emph{Performance Analysis}}:
  Let $M^{(b)}_j$ and $\hat{M}^{(b)}_j$ denote the random encoded and decoded indices for $\bm{S}^{(b)}_j$. 
  We first define the events $\mathcal{E}_1^{(b)}$, $b=1, 2, \dots, B+1$, in \eqref{eq:errorevent} for terminal~1. 
  We analogously define the events $\mathcal{E}_2^{(b)}$ for terminal~$2$ (not shown here) and consider the error event $\mathcal{E}=\cup_{b=1}^{B+1} \mathcal{E}_1^{(b)}\cup \mathcal{E}_2^{(b)}$. 
  The expected distortion of terminal~$j$'s source reconstruction (averaged with respect to all codebooks, source messages, channel inputs, and channel outputs) can be bounded by
  \begin{IEEEeqnarray}{rCl}
  \frac{1}{B}\sum_{b=1}^B\mathbb{E}[d_j(\bm{S}^{(b)}_j, \hat{\bm{S}}^{(b)}_j)] &\le&  \Pr(\mathcal{E}) d_{j, \max}+\frac{1}{B}\sum_{b=1}^B \Pr\big(\overline{\mathcal{E}}\big)\mathbb{E}[d_j(\bm{S}^{(b)}_j, \hat{\bm{S}}^{(b)}_j)|\overline{\mathcal{E}}]\label{eq:distanalysisCE}\IEEEeqnarraynumspace\\
  &\le&  \Pr(\mathcal{E}) d_{j, \max}+\frac{1}{B}\sum_{b=1}^B (1+\epsilon)\mathbb{E}[d_j(S^{(b)}_j, \hat{S}^{(b)}_j)]\label{eq:distanalysisavertypical}\\
  &=&  \Pr(\mathcal{E}) d_{j, \max}+(1+\epsilon)\mathbb{E}[d_j(S_j, \hat{S}_j)]\label{eq:distanalysisstat}\\
  &\le&  \Pr(\mathcal{E}) d_{j, \max}+(1+\epsilon) D_j,
  \end{IEEEeqnarray}
  where \eqref{eq:distanalysisCE} follows from $\mathbb{E}[d_j(\bm{S}^{(b)}_j, \hat{\bm{S}}^{(b)}_j)|\mathcal{E}]\le d_{j, \max}$ with $d_{j, \max}\triangleq \max_{s_j, \hat{s}_j}d_j(s_j, \hat{s}_j)$, \eqref{eq:distanalysisavertypical} is due to the typical average lemma \cite{kim2011}, \eqref{eq:distanalysisstat} follows from the stationarity of the Markov chain, and the last inequality holds by assumption. 
  
  If we can further show that $\Pr\big(\mathcal{E}\big)\to 0$ and the joint source-channel coding rate goes to one as both $n$ and $B$ go to infinity, then the distortion pair $((1+\epsilon)D_1, (1+\epsilon)D_2)$ is achievable. 
  Note that it suffices to show that $\Pr\big(\mathcal{E}^{(1)}_j\big)\to 0$ and $\Pr\big(\mathcal{E}^{(b)}_j\cap\overline{\mathcal{E}}_j^{(b-1)}\big)\to 0$ for all $j=1, 2$ and $b=2, 3, \dots, B+1$ since by the identity $\cup_{b=1}^{B}\mathcal{E}_j^{(b)}=\mathcal{E}_j^{(1)}\cup\big(\cup_{b=2}^{B}\mathcal{E}_j^{(b)}\cap\overline{\mathcal{E}}_j^{(b-1)}\big)$, we have
  \begin{IEEEeqnarray}{l}
    \scalemath{1}{\Pr(\mathcal{E})\le \Pr(\mathcal{E}_1^{(1)})+\Pr(\mathcal{E}_2^{(1)})}\scalemath{1}{+\sum_{b=2}^{B+1}\left(\Pr(\mathcal{E}_1^{(b)}\cap\overline{\mathcal{E}}_1^{(b-1)})+\Pr(\mathcal{E}_2^{(b)}\cap\overline{\mathcal{E}}_2^{(b-1)})\right)}.\nonumber
  \end{IEEEeqnarray}
  
  Due to symmetry, we only analyze $\Pr\big(\mathcal{E}^{(1)}_1\big)$ and $\Pr\big(\mathcal{E}^{(b)}_1\cap\overline{\mathcal{E}}_1^{(b-1)}\big)$.
  For $j=1, 2$ and $b=1, 2, \dots, B+1$, we first define
  \begin{IEEEeqnarray}{l}
  \scalemath{0.93}{\mathcal{F}^{(b)}_j =\{(\bm{S}_j^{(b)}, \bm{U}_j^{(b)}(m_j^{(b)}))\notin\mathcal{T}^{(n)}_{\epsilon_1}\ \text{for all}\ m_j^{(b)}\}},\nonumber\\
  \scalemath{0.93}{\mathcal{F}^{(b)}_3=\{(\bm{S}_1^{(b)}, \bm{S}_2^{(b)}, \bm{U}_1^{(b)}(M_1^{(b)}), \bm{U}_2^{(b)}(M_2^{(b)}), \tilde{\bm{S}}_1^{(b)}, \tilde{\bm{S}}_2^{(b)}}, \nonumber\\
  \qquad\qquad\qquad \ \ \scalemath{0.93}{\tilde{\bm{U}}_1^{(b)}(M_1^{(b-1)}), \tilde{\bm{U}}_2^{(b)}(M_2^{(b-1)}), \tilde{\bm{W}}_1^{(b)}, \tilde{\bm{W}}_2^{(b)}\bm{X}_1^{(b)}, \bm{X}_2^{(b)}, \bm{Y}_1^{(b)}, \bm{Y}_2^{(b)})\notin\mathcal{T}^{(n)}_{\epsilon}\}},\nonumber\\
  \scalemath{0.93}{\mathcal{F}^{(b)}_4 =\{\exists\ \hat{m}^{(b-1)}_1\neq M^{(b-1)}_1\ \text{s.t.}\ (\bm{S}_2^{(b)}, \bm{U}_2^{(b)}(M_2^{(b)}), \tilde{\bm{S}}_2^{(b)}}, \nonumber\\  
  \qquad\qquad\qquad \ \ \scalemath{0.93}{\tilde{\bm{U}}_1^{(b)}(\hat{m}_1^{(b-1)}), \tilde{\bm{U}}_2^{(b)}(M_2^{(b-1)}), \tilde{\bm{W}}_2^{(b)}, \bm{X}_2^{(b)}, \bm{Y}_2^{(b)})\in\mathcal{T}^{(n)}_{\epsilon}\}},\nonumber
  \end{IEEEeqnarray}
  with the exception that $\mathcal{F}^{(1)}_3\triangleq\mathcal{E}_1^{(1)}$ and $\mathcal{F}^{(B+1)}_3\triangleq\mathcal{E}_1^{(B+1)}$ due to the initialization and termination phases of the encoding process. 
  We will use the following results to obtain \eqref{eq:mainconda}; detailed proofs of the claims are given in the next section. \medskip
  
  \noindent {\it Claim~1}: For $b=2, 3, \dots, B+1$, the event $\overline{\mathcal{F}}^{(b)}_3\cap\overline{\mathcal{F}}^{(b)}_4$ implies that $\hat{M}_1^{(b-1)}=M_1^{(b-1)}$.\smallskip
  
  \noindent {\it Claim~2}: $\mathcal{E}_1^{(1)}\subseteq \mathcal{F}^{(1)}_1\cup \mathcal{F}^{(1)}_2\cup (\overline{\mathcal{F}}^{(1)}_1\cap\overline{\mathcal{F}}^{(1)}_2\cap\mathcal{E}_1^{(1)})$ \smallskip
  
  \noindent {\it Claim~3}: The inclusion $\mathcal{E}_1^{(b)}\cap\overline{\mathcal{E}}_1^{(b-1)}\subseteq \mathcal{F}^{(b)}_1\cup \mathcal{F}^{(b)}_2\cup (\overline{\mathcal{F}}^{(1)}_1\cap\overline{\mathcal{F}}^{(1)}_2\cap\mathcal{F}_3^{(b)}\cap\overline{\mathcal{E}}_1^{(b-1)})\cup \mathcal{F}_4^{(b)}$ holds for $b=2, 3, \dots, B$.\smallskip

  \noindent {\it Claim~4}: $\mathcal{E}_1^{(B+1)}\cap\overline{\mathcal{E}}_1^{(B)}\subseteq (\mathcal{F}_3^{(B+1)}\cap\overline{\mathcal{E}}_1^{(B)}) \cup \mathcal{F}_4^{(B+1)}$\smallskip
  
  \noindent {\it Claim~5}: If $R^{(1)}_j>I(S_j; U_j)+\delta_1(\epsilon_1)$, then $\lim_{n\to\infty}\Pr\big(\mathcal{E}_j^{(1)}\big)=0$.\smallskip
  
  \noindent {\it Claim~6}: If $R^{(B)}_1<I(\tilde{U}_1; S_{2}, U_{2}, \tilde{S}_{2}, \tilde{U}_{2}, \tilde{W}_{2}, X_{2}, Y_{2})-\delta(\epsilon)$, then $\lim_{n\to\infty}\Pr\big(\mathcal{E}_1^{(B+1)}\cap\overline{\mathcal{E}}_1^{(B)}\big)=0$.\smallskip
  
  \noindent {\it Claim~7}: For $b=2, 3, \dots, B$, if $R^{(b)}_j > I(S_j; U_j)+\delta_1(\epsilon_1)$ and $R^{(b-1)}_1<I(\tilde{U}_1; S_{2}, U_{2}, \allowbreak\tilde{S}_{2}, \allowbreak\tilde{U}_{2}, \allowbreak\tilde{W}_{2}, X_{2}, Y_{2})-\delta(\epsilon)$, then $\lim_{n\to\infty}\Pr\big(\mathcal{E}_{1}^{(b)}\cap\overline{\mathcal{E}}_{1}^{(b-1)}\big)=0$.\medskip
  
  The non-negative quantities $\delta_1(\epsilon_1)$ and $\delta(\epsilon)$ above arise from the standard typicality arguments and $\lim_{\epsilon_1\to 0}\delta_1(\epsilon_1)=0$ and $\lim_{\epsilon\to 0}\delta(\epsilon)=0$. 
  Swapping the role of terminals~1 and~2, we obtain $\lim_{n\to\infty}\Pr\big(\mathcal{E}_2^{(1)}\big)=0$ and are such that $\lim_{n\to\infty}\Pr\big(\mathcal{E}_{2}^{(b)}\cap\overline{\mathcal{E}}_{2}^{(b-1)}\big)=0$ for $b=2, 3, \dots, B+1$ provided that $R^{(b)}_j>I(S_j; U_j)+\delta_1(\epsilon_1)$
  for $j=1, 2$ and $b=1, 2, \dots, B$ and $R^{(b-1)}_2<\allowbreak I(\tilde{U}_2; S_{1}, U_{1}, \allowbreak\tilde{S}_{1}, \allowbreak\tilde{U}_{1}, \allowbreak\tilde{W}_{1}, \allowbreak X_{1}, \allowbreak Y_{1})-\delta(\epsilon)$
  for $b=2, 3, \dots, B+1$. 
  Combining all conditions above then gives the two inequalities in \eqref{eq:mainconds}. 
  To complete the proof, we first increase $B$ so that the JSCC rate $B/(B+1)$ is close to one. 
  Fixing this choice of $B$, we next make $n$ sufficiently large to ensure that all joint typicality requirements behind Claims~5-7 (and similar claims for terminal~2) are satisfied. 
  As now we have $\lim_{n\to\infty}\Pr(\mathcal{E}){=}0$ (provided that all conditions hold) and $\epsilon$ is arbitrary, the distortion pair $(D_1, D_2)$ is achievable. \hfill \IEEEQEDhere

  \subsection{Auxiliary Results for the Proof of Theorem~\ref{thm:main}}\label{subsec:claims}
  Here, we prove Claims~1-7 in the proof of Theorem~\ref{thm:main}. 
  We use $\mathcal{T}_{\epsilon}^{(n)}(\cdot|\cdot)$ to denote conditional typical sets. 

  \noindent \textbf{Claim~1}: For $b=2, 3, \dots, B+1$, the event $\overline{\mathcal{F}}^{(b)}_3\cap\overline{\mathcal{F}}^{(b)}_4$ implies that $\hat{M}_1^{(b-1)}=M_1^{(b-1)}$.
  \begin{IEEEproof}
    $\overline{\mathcal{F}}_3^{(b)}$ implies that 
    \[(\bm{S}_2^{(b)}, \bm{U}_2^{(b)}, \allowbreak\tilde{\bm{S}}_2^{(b)}, \allowbreak\tilde{\bm{U}}_1^{(b)}(M^{(b-1)}_1), \allowbreak\tilde{\bm{U}}_2^{(b)}(M^{(b-1)}_2), \allowbreak\tilde{\bm{W}}_2^{(b)},\allowbreak \bm{X}_2^{(b)}, \bm{Y}_2^{(b)})\in\mathcal{T}_{\epsilon}^{(n)}.\] 
    Thus, we have that $\hat{M}^{(b-1)}_1=M^{(b-1)}_1$ under $\overline{\mathcal{F}}_3^{(b)}\cap \overline{\mathcal{F}}_4^{(b)}$.
  \end{IEEEproof}\medskip
  
  \noindent \textbf{Claim~2}: $\mathcal{E}_1^{(1)}\subseteq \mathcal{F}^{(1)}_1\cup \mathcal{F}^{(1)}_2\cup (\overline{\mathcal{F}}^{(1)}_1\cap\overline{\mathcal{F}}^{(1)}_2\cap\mathcal{E}_1^{(1)})$
  \begin{IEEEproof}
    This follows since the right-hand-side is equal to $\mathcal{E}_1^{(1)}\cup\mathcal{F}^{(1)}_1\cup \mathcal{F}^{(1)}_2$.
  \end{IEEEproof}\medskip
  
  \noindent \textbf{Claim~3}: The inclusion $\mathcal{E}_1^{(b)}\cap\overline{\mathcal{E}}_1^{(b-1)}\subseteq \mathcal{F}^{(b)}_1\cup \mathcal{F}^{(b)}_2\cup (\overline{\mathcal{F}}^{(b)}_1\cap\overline{\mathcal{F}}^{(b)}_2\cap\mathcal{F}_3^{(b)}\cap\overline{\mathcal{E}}_1^{(b-1)})\cup \mathcal{F}_4^{(b)}$ holds for $b=2, 3, \dots, B$. 
  \begin{IEEEproof}
    Claim~1 implies that $\overline{\mathcal{F}}_3^{(b)}\cap \overline{\mathcal{F}}_4^{(b)}\subseteq\overline{\mathcal{E}}_1^{(b)}$ and hence $\mathcal{E}_1^{(b)}\subseteq\mathcal{F}_3^{(b)}\cup\mathcal{F}_4^{(b)}$.
    Together with the facts that \[\mathcal{E}_1^{(b)}\cap\overline{\mathcal{E}}_1^{(b-1)}\subseteq(\mathcal{F}_3^{(b)}\cap\overline{\mathcal{E}}_1^{(b-1)})\cup(\mathcal{F}_4^{(b)}\cap\overline{\mathcal{E}}_1^{(b-1)})\subseteq (\mathcal{F}_3^{(b)}\cap\overline{\mathcal{E}}_1^{(b-1)})\cup\mathcal{F}_4^{(b)}\] and that 
    \begin{IEEEeqnarray}{rCl}
      \mathcal{F}_3^{(b)}\cap\overline{\mathcal{E}}_1^{(b-1)}&=& (\mathcal{F}_3^{(b)}\cap\overline{\mathcal{E}}_1^{(b-1)}\cap(\mathcal{F}^{(b)}_1\cup\mathcal{F}^{(b)}_2))\cup(\mathcal{F}_3^{(b)}\cap\overline{\mathcal{E}}_1^{(b-1)}\cap\overline{\mathcal{F}^{(b)}_1\cup\mathcal{F}^{(b)}_2})\nonumber\\
      &\subseteq& \mathcal{F}^{(b)}_1\cup \mathcal{F}^{(b)}_2\cup (\overline{\mathcal{F}^{(b)}_1\cup\mathcal{F}^{(b)}_2}\cap\mathcal{F}_3^{(b)}\cap\overline{\mathcal{E}}_1^{(b-1)}),\nonumber
    \end{IEEEeqnarray}
    we obtain the desired inclusion relationship.
  \end{IEEEproof}\medskip

  \noindent \textbf{Claim~4}: $\mathcal{E}_1^{(B+1)}\cap\overline{\mathcal{E}}_1^{(B)}\subseteq (\mathcal{F}_3^{(B+1)}\cap\overline{\mathcal{E}}_1^{(B)}) \cup \mathcal{F}_4^{(B+1)}$
  \begin{IEEEproof}
    The result follows from the proof of Claim~3. 
  \end{IEEEproof}
  \medskip
  
  \noindent \textbf{Claim~5}: If $R^{(1)}_j>I(S_j; U_j)+\delta_1(\epsilon_1)$, then $\lim_{n\to\infty}\Pr\big(\mathcal{E}_j^{(1)}\big)=0$.
  \begin{IEEEproof}
    Due to Claim~2, it suffices to show that $\lim_{n\to\infty}\Pr(\mathcal{F}^{(1)}_j)=0$ for $j=1, 2$ and $\lim_{n\to\infty}\Pr(\overline{\mathcal{F}^{(1)}_1\cup\mathcal{F}^{(1)}_2}\cap\mathcal{E}_1^{(1)})=0$ under the hypothesis. 
    For $\Pr(\mathcal{F}^{(1)}_j)$, we define a non-typical set $\mathcal{A}_j=\{\bm{S}^{(1)}_j\notin\mathcal{T}^{(n)}_{\epsilon_0}\}$ for some $\epsilon_0< \epsilon_1$, $j=1, 2$. Then, $\mathcal{F}^{(1)}_j\subseteq\mathcal{A}_j\cup (\mathcal{F}^{(1)}_j\cap\overline{\mathcal{A}}_j)$.
    Clearly, $\lim_{n\to\infty}\Pr(\mathcal{A}_j)=0$ due to the weak law of large numbers, and $\Pr(\mathcal{F}^{(1)}_j\cap\overline{\mathcal{A}}_j)\le\Pr(\mathcal{F}^{(1)}_j|\overline{\mathcal{A}}_j)$. For $\Pr(\mathcal{F}^{(1)}_j|\overline{\mathcal{A}}_j)$, we apply the covering lemma \cite[Lemma~3.3]{kim2011} with the correspondences \[X\leftrightarrow\emptyset, U\leftrightarrow S_j, \hat{X}\leftrightarrow U_j, R\leftrightarrow R^{(1)}_j, \epsilon'\leftrightarrow\epsilon_0, \text{\ and\ } \epsilon\leftrightarrow\epsilon_1\] to obtain that if $R^{(1)}_j>I(S_j; U_j)+\delta(\epsilon_1)$, then $\lim_{n\to\infty}\Pr(\mathcal{F}^{(1)}_j|\overline{\mathcal{A}}_j)=0$.
    Thus, we obtain $\lim_{n\to\infty}\Pr(\mathcal{F}^{(1)}_j)=0$ under the hypothesis for $j=1, 2$.
    
    The proof of $\lim_{n\to\infty}\Pr(\overline{\mathcal{F}^{(1)}_1\cup\mathcal{F}^{(1)}_2}\cap\mathcal{E}_j^{(1)})=0$ is more involved. 
    For $\epsilon_2$ and $\epsilon_3$ such that $\epsilon_1<\epsilon_2<\epsilon_3$, let 
    \[\mathcal{B}_1\triangleq\{(\bm{S}^{(1)}_1, \bm{S}^{(1)}_2, \bm{U}^{(1)}_1(M_1^{(1)}))\notin\mathcal{T}^{(n)}_{\epsilon_2}\}\] 
    and 
    \[\mathcal{B}_2\triangleq\{(\bm{S}^{(1)}_1, \bm{S}^{(1)}_2, \allowbreak\bm{U}^{(1)}_1(M_1^{(1)}), \allowbreak\bm{U}^{(1)}_2(M_2^{(1)}))\notin\mathcal{T}^{(n)}_{\epsilon_3}\}.\] 
    We first show that conditional on the event $\overline{\mathcal{F}^{(1)}_1\cup\mathcal{F}^{(1)}_2}$, we have that $\lim_{n\to\infty}\Pr(\mathcal{B}_2)=0$.
    We begin by considering the inclusion relationship: \[\mathcal{B}_2\subseteq \mathcal{F}_1^{(1)} \cup (\mathcal{B}_1\cap\overline{\mathcal{F}}_1^{(1)}) \cup \mathcal{F}_2^{(1)} \cup (\mathcal{B}_2\cap \overline{\mathcal{B}}_1\cap \overline{\mathcal{F}}_2^{(1)}).\]  
    Using union bound, we have that 
    \begin{IEEEeqnarray}{rCl}
        \Pr(\mathcal{B}_2)& \le &  \Pr(\mathcal{F}_1^{(1)}) + \Pr(\mathcal{F}_2^{(1)}) + \Pr(\mathcal{B}_1\cap\overline{\mathcal{F}}_1^{(1)}) +\Pr(\mathcal{B}_2\cap \overline{\mathcal{B}}_1\cap \overline{\mathcal{F}}_2^{(1)})\nonumber\\
        & \le &  \Pr(\mathcal{F}_1^{(1)}) + \Pr(\mathcal{F}_2^{(1)}) + \Pr(\mathcal{B}_1|\overline{\mathcal{F}}_1^{(1)}) + \Pr(\mathcal{B}_2|\overline{\mathcal{B}}_1\cap \overline{\mathcal{F}}_2^{(1)})\label{eq:xx1}.
    \end{IEEEeqnarray}
    Now, applying the conditional typicality lemma \cite[Section~2.5]{kim2011} with the correspondences 
    \[X\leftrightarrow (S_1, U_1), Y\leftrightarrow S_2, \epsilon'\leftrightarrow\epsilon_1, \text{\ and\ }\epsilon\leftrightarrow\epsilon_2,\] we have that $\lim_{n\to\infty} \Pr(\overline{\mathcal{B}}_1|\overline{\mathcal{F}}_1^{(1)})=~1$. 
    Similarly, applying the conditional typical lemma with the correspondences: \[X\leftrightarrow (S_1, S_2, U_1), Y\leftrightarrow U_2, \epsilon'\leftrightarrow\epsilon_2, \text{\ and\ } \epsilon\leftrightarrow\epsilon_3,\] one further obtains that $\lim_{n\to\infty} \Pr(\overline{\mathcal{B}}_2|\overline{\mathcal{B}}_1\cap \overline{\mathcal{F}}_2^{(1)})=1$. 
    Together with the first part of the proof and \eqref{eq:xx1}, we conclude that $\lim_{n\to\infty}\Pr(\mathcal{B}_2)=0$. 

    We next use the inclusion $\overline{\mathcal{F}^{(1)}_1\cup\mathcal{F}^{(1)}_2}\cap\mathcal{E}_j^{(1)}\subseteq \mathcal{B}_2\cup\mathcal{E}_j^{(1)}\subseteq\mathcal{B}_2\cup (\mathcal{E}_j^{(1)}\cap\overline{\mathcal{B}}_2)$, which yields the inequality $\Pr(\overline{\mathcal{F}^{(1)}_1\cup\mathcal{F}^{(1)}_2}\cap\mathcal{E}_j^{(1)})\le \Pr(\mathcal{B}_2)+ \Pr(\mathcal{E}_j^{(1)}|\overline{\mathcal{B}}_2)$. 
    For $\Pr(\mathcal{E}_j^{(1)}|\overline{\mathcal{B}}_2)$, since $(\tilde{\bm{S}}^{(1)}_1, \tilde{\bm{S}}^{(1)}_2, \tilde{\bm{U}}^{(1)}_1, \allowbreak\tilde{\bm{U}}^{(1)}_2, \allowbreak\tilde{\bm{W}}^{(1)}_1, \allowbreak\tilde{\bm{W}}^{(1)}_2)$ is generated according to \eqref{eq:cwgen1} (and is independent of $(\bm{S}^{(1)}_1, \bm{S}^{(1)}_2, \bm{U}^{(1)}_1, \bm{U}^{(1)}_2)$) and the channel input $\bm{X}_1^{(1)}$ is generated component-wise, the conditional typicality lemma implies that \vspace{-0.1cm}
    \[\lim_{n\to\infty}\Pr(\bm{S}^{(1)}_1, \bm{S}^{(1)}_2, \bm{U}^{(1)}_1, \allowbreak\bm{U}^{(1)}_2, \tilde{\bm{S}}^{(1)}_1, \tilde{\bm{S}}^{(1)}_2, \tilde{\bm{U}}^{(1)}_1, \tilde{\bm{U}}^{(1)}_2, \tilde{\bm{W}}^{(1)}_1, \tilde{\bm{W}}^{(1)}_2, \bm{X}_1^{(1)}, \allowbreak\bm{X}_2^{(1)})\in\mathcal{T}_{\epsilon_4}^{(n)})=1\] under $\overline{\mathcal{B}}_2$ for some $\epsilon_4>\epsilon_3$.
    Applying the conditional typicality lemma again with the correspondences 
    \[X\leftrightarrow (S_1, S_2, U_1, U_2, \tilde{S}_1, \tilde{S}_2, \allowbreak\tilde{U}_1, \allowbreak\tilde{U}_2, \allowbreak\tilde{W}_1, \tilde{W}_2, \tilde{X}_1, \allowbreak\tilde{X}_2), Y\leftrightarrow (Y_1, Y_2), \epsilon'\leftrightarrow \epsilon_4, \text{\ and\ } \epsilon\leftrightarrow \epsilon,\] and using the memoryless property of the channel, we further have that $\lim_{n\to\infty}\Pr(\mathcal{E}_j^{(1)}|\overline{\mathcal{B}}_2)=0$.  Combining this with \eqref{eq:xx1} implies $\lim_{n\to\infty}\Pr(\overline{\mathcal{F}^{(1)}_1\cup\mathcal{F}^{(1)}_2}\cap\mathcal{E}_j^{(1)})=0$, which completes the proof of the claim. 
\end{IEEEproof}
\medskip
  
\noindent \textbf{Claim~6}: If $R^{(B)}_1<I(\tilde{U}_1; S_{2}, U_{2}, \tilde{S}_{2}, \tilde{U}_{2}, \tilde{W}_{2}, X_{2}, Y_{2})-\delta(\epsilon)$, then $\lim_{n\to\infty}\Pr\big(\mathcal{E}_1^{(B+1)}\cap\overline{\mathcal{E}}_1^{(B)}\big)=0$.
\begin{IEEEproof}
  With the help of Claim~4, it suffices to show that $\lim_{n\to\infty}\allowbreak\Pr(\mathcal{F}_3^{(B+1)}\cap\overline{\mathcal{E}}_1^{(B)})=0$ and $\lim_{n\to\infty}\Pr(\mathcal{F}_4^{(B+1)})=0$ under the hypothesis. 
  To obtain the first result, we follow the proof of Claim~5.   
  Consider the inequality $\Pr(\mathcal{F}_3^{(B+1)}\cap\overline{\mathcal{E}}_1^{(B)})\le \Pr(\mathcal{F}_3^{(B+1)}|\overline{\mathcal{E}}_1^{(B)})$. 
  Conditioning on $\overline{\mathcal{E}}_1^{(B)}$ clearly imposes a joint typicality constraint on the sequence $(\tilde{\bm{S}}_1^{(B+1)}, \allowbreak\tilde{\bm{S}}_2^{(B+1)}, \allowbreak\tilde{\bm{U}}_1^{(B+1)}, \allowbreak\tilde{\bm{U}}_2^{(B+1)}, \allowbreak\tilde{\bm{W}}_1^{(B+1)}, \allowbreak\tilde{\bm{W}}_2^{(B+1)})$ in the event $\mathcal{F}_3^{(B+1)}$. 
  We also know that the sequence $(\bm{S}^{(B+1)}_1, \allowbreak\bm{S}^{(B+1)}_2, \allowbreak\bm{U}^{(B+1)}_1, \allowbreak\bm{U}^{(B+1)}_2)$ in the event $\mathcal{F}_3^{(B+1)}$ will be jointly typical with high probability due to \eqref{eq:cwgen2} and the weak law of large numbers.
  Using these observations, we apply the conditional typicality lemma twice, as in the last part of the proof of Claim~5, to conclude that $\lim_{n\to\infty}\allowbreak\Pr(\mathcal{F}_3^{(B+1)}\cap\overline{\mathcal{E}}_1^{(B)})=0$.

  To analyze $\Pr(\mathcal{F}_4^{(B+1)})$, we may assume that $(M^{(B)}_{1}, M^{(B)}_{2})=(1, 1)\triangleq \bm{M}^{(B)}_{1, 1}$ by the symmetry of random codebook generation and the encoding procedure. 
  Then, we have two Markov chain relationships for $m_1\neq 1$:
  \begin{IEEEeqnarray}{l}
      \tilde{\bm{U}}_1^{(B+1)}(m_1)\markov (\bm{S}^{(B+1)}_1, \bm{S}^{(B+1)}_2, \bm{U}^{(B+1)}_1, \bm{U}^{(B+1)}_2, \tilde{\bm{S}}^{(B+1)}_1, \tilde{\bm{S}}^{(B+1)}_2, \tilde{\bm{U}}^{(B+1)}_1(1), \tilde{\bm{U}}^{(B+1)}_2(1), \nonumber\\
      \quad\qquad\qquad\qquad\qquad\ \tilde{\bm{W}}^{(B+1)}_1, \tilde{\bm{W}}^{(B+1)}_2)\markov (\bm{X}^{(B+1)}_1, \bm{X}^{(B+1)}_2)\markov (\bm{Y}^{(B+1)}_1, \bm{Y}^{(B+1)}_2) \label{eq:mc1}\IEEEeqnarraynumspace   
  \end{IEEEeqnarray}
  and 
  \begin{IEEEeqnarray}{l}
      \tilde{\bm{U}}_1^{(B+1)}(m_1)\markov (\tilde{\bm{S}}^{(B+1)}_1, \tilde{\bm{U}}^{(B+1)}_1(1)) \markov (\bm{S}^{(B+1)}_1, \bm{S}^{(B+1)}_2, \bm{U}^{(B+1)}_1, \bm{U}^{(B+1)}_2, \tilde{\bm{S}}^{(B+1)}_2, \nonumber\IEEEeqnarraynumspace\\
      \quad\qquad\qquad\qquad\qquad\qquad\qquad\qquad\quad\ \  \tilde{\bm{U}}^{(B+1)}_2(1), \tilde{\bm{W}}^{(B+1)}_1, \tilde{\bm{W}}^{(B+1)}_2, \bm{X}^{(B+1)}_1, \bm{X}^{(B+1)}_2).\label{eq:mc2}\IEEEeqnarraynumspace   
  \end{IEEEeqnarray}

  To simplify the derivation, we define 
  \[\bm{A}_1(\hat{m}_1^{(B)})=(\bm{S}_2^{(B+1)}, \bm{U}_2^{(B+1)}, \tilde{\bm{S}}_2^{(B+1)}, \tilde{\bm{U}}_1^{(B+1)}(\hat{m}_1^{(B)}), \tilde{\bm{U}}_2^{(B+1)}(1), \tilde{\bm{W}}_2^{(B+1)}, \allowbreak\bm{X}_2^{(B+1)}, \bm{Y}_2^{(B+1)})\] 
  and let $\bm{a}_1=(\bm{s}_2, \bm{u}_2, \tilde{\bm{s}}_2, \tilde{\bm{u}}_1, \tilde{\bm{u}}_2, \tilde{\bm{w}}_2, \allowbreak\bm{x}_2, \bm{y}_2)$ to denote a realization of $\bm{A}_1(\hat{m}_1^{(B)})$. 
  When excluding the variable $\tilde{\bm{U}}_1^{(B+1)}(\hat{m}_1^{(B)})$ (resp., $(\tilde{\bm{U}}_1^{(B+1)}(\hat{m}_1^{(B)}), \bm{X}_2^{(B+1)}, \bm{Y}_2^{(B+1)})$) from $\bm{A}_1(\hat{m}_1^{(B)})$, we let the remaining tuples denoted as $\bm{A}_2$ (resp., $\bm{A}_3$). 
  Note that when $\bm{a}_1$ is given, $\bm{a}_2$ and $\bm{a}_3$ are determined as well. 
  Moreover, we define \[\bm{B}=(\bm{S}_1^{(B+1)}, \bm{U}_1^{(B+1)}, \tilde{\bm{S}}_1^{(B+1)}, \tilde{\bm{U}}_1^{(B+1)}(1), \tilde{\bm{W}_1}^{(B+1)})\] and let $\bm{b}=(\bm{s}_1, \bm{u}'_1, \tilde{\bm{s}}_1, \tilde{\bm{u}}'_1, \bm{w}_1)$ to denote a realization of it. 
  In the following, we find an upper bound for $\Pr(\mathcal{F}_4^{(B+1)})$ using the fact that $\Pr(\mathcal{F}_4^{(B+1)})=\Pr(\mathcal{F}_4^{(B+1)}|\bm{M}^{(B)}_{1, 1})$: 
  \begin{IEEEeqnarray}{l}
  \Pr(\mathcal{F}_4^{(B+1)}|\bm{M}^{(B)}_{1, 1}) \nonumber\\
  \ \ \le \sum\limits_{\hat{m}_1=2}^{2^{\mli{nR}_1^{(B)}}}\sum_{\bm{a}_1\in\mathcal{T}^{(n)}_{\epsilon}}\Pr(\bm{A}_1(\hat{m}_1)=\bm{a}_1|\bm{M}^{(B)}_{1, 1})\label{eq:e4s1}\\
  \ \ = \sum\limits_{\hat{m}_1=2}^{2^{\mli{nR_1^{(B)}}}}\sum_{\bm{a}_1\in\mathcal{T}^{(n)}_{\epsilon}}\sum_{\bm{b}}\Pr(\bm{A}_1(\hat{m}_1)=\bm{a}_1, \bm{B}=\bm{b}|\bm{M}^{(B)}_{1, 1})\label{eq:e4s2}\IEEEeqnarraynumspace\\
  \ \ = \sum\limits_{\hat{m}_1=2}^{2^{\mli{nR}_1^{(B)}}}\sum_{\bm{a}_1\in\mathcal{T}^{(n)}_{\epsilon}}\sum_{\bm{b}}\Pr(\bm{X}^{(B+1)}_2=\bm{x}_2, \bm{Y}^{(B+1)}_2=\bm{y}_2|\bm{M}^{(B)}_{1, 1})\nonumber\\
  \qquad\qquad \Pr(\bm{A}_3=\bm{a}_3, \bm{B}=\bm{b}|\bm{X}^{(B+1)}_2=\bm{x}_2, \bm{Y}^{(B+1)}_2=\bm{y}_2, \bm{M}^{(B)}_{1, 1})\nonumber\\
  \qquad\qquad\qquad\qquad\qquad \Pr(\tilde{\bm{U}}_1^{(B+1)}(\hat{m}_1)=\tilde{\bm{u}}_1|\bm{A}_2=\bm{a}_2, \bm{B}=\bm{b}, \bm{M}^{(B)}_{1, 1})\IEEEeqnarraynumspace\label{eq:e4s3}\\
  \ \ = \sum\limits_{\hat{m}_1=2}^{2^{\mli{nR}_1^{(B)}}}\sum_{\bm{a}_1\in\mathcal{T}^{(n)}_{\epsilon}}\sum_{\bm{b}}\Pr(\bm{X}^{(B+1)}_2=\bm{x}_2, \bm{Y}^{(B+1)}_2=\bm{y}_2|\bm{M}^{(B)}_{1, 1})\nonumber\\
  \qquad\qquad \Pr(\bm{A}_3=\bm{a}_3, \bm{B}=\bm{b}|\bm{X}^{(B+1)}_2=\bm{x}_2, \bm{Y}^{(B+1)}_2=\bm{y}_2, \bm{M}^{(B)}_{1, 1})\nonumber\IEEEeqnarraynumspace\\
  \qquad\qquad\qquad\qquad\qquad \Pr(\tilde{\bm{U}}_1^{(B+1)}(\hat{m}_1)=\tilde{\bm{u}}_1|\tilde{\bm{S}}_1^{(B+1)}=\tilde{\bm{s}}_1, \tilde{\bm{U}}_1^{(B+1)}(1)=\bm{u}'_1, M_1^{(B)}=1)\IEEEeqnarraynumspace\label{eq:e4s4}\\
  \ \ \le \sum\limits_{\hat{m}_1=2}^{2^{\mli{nR}_1^{(B)}}}\sum_{\bm{a}_1\in\mathcal{T}^{(n)}_{\epsilon}}\Pr(\bm{X}^{(B+1)}_2=\bm{x}_2, \bm{Y}^{(B+1)}_2=\bm{y}_2|\bm{M}^{(B)}_{1, 1})\cdot (1+\epsilon)\prod_{i=1}^n P_{\tilde{U}^{(B+1)}_1}(\tilde{u}_{1,i})\nonumber\\
  \qquad\qquad \sum_{\bm{b}}\Pr(\bm{A}_3=\bm{a}_3, \bm{B}=\bm{b}|\bm{X}^{(B+1)}_2=\bm{x}_2, \bm{Y}^{(B+1)}_2=\bm{y}_2, \bm{M}^{(B)}_{1, 1})\label{eq:e4s5}\IEEEeqnarraynumspace\\
  \ \ = (1+\epsilon)\sum\limits_{\hat{m}_1=2}^{2^{\mli{nR}_1^{(B)}}}\sum_{\bm{a}_1\in\mathcal{T}^{(n)}_{\epsilon}}\Pr(\bm{A}_2=\bm{a}_2|\bm{M}^{(B)}_{1, 1})\prod_{i=1}^n P_{\tilde{U}^{(B+1)}_1}(\tilde{u}_{1,i})\label{eq:e4s6}\\
  \ \ \le (1+\epsilon)\cdot 2^{\mli{nR}_1^{(B)}}\sum_{\bm{a}_2\in\mathcal{T}^{(n)}_{\epsilon}}\sum_{\tilde{\bm{u}}_1\in\mathcal{T}^{(n)}_\epsilon(\tilde{U}_1|\bm{a}_2)}\Pr(\bm{A}_2=\bm{a}_2|\bm{M}^{(B)}_{1, 1})\prod_{i=1}^n P_{\tilde{U}_1}(\tilde{u}^{(B+1)}_{1,i})\\
  \ \ \le (1+\epsilon)\cdot 2^{\mli{nR}_1^{(B)}}\sum_{\bm{a}_2\in\mathcal{T}^{(n)}_{\epsilon}}|\mathcal{T}^{(n)}_\epsilon(\tilde{U}_1|\bm{a}_2)|\cdot\Pr(\bm{A}_2=\bm{a}_2|\bm{M}^{(B)}_{1, 1})\cdot 2^{-n(H(\tilde{U}_1)-\delta_1(\epsilon))}\label{eq:e4s7}\\
  \ \ \le (1+\epsilon)\cdot 2^{\mli{nR}_1^{(B)}}2^{n(H(\tilde{U}_1|S_2, U_2, \tilde{S}_2, \tilde{U}_2, \tilde{W}_2, X_2, Y_2)+\delta_2(\epsilon))}\cdot 2^{-n(H(\tilde{U}_1)-\delta_1(\epsilon))}\label{eq:e4s8}\\
  \ \ \le (1+\epsilon)\cdot 2^{n(R_1^{(B)}-I(\tilde{U}_1; S_2, U_2, \tilde{S}_2, \tilde{U}_2, \tilde{W}_2, X_2, Y_2)+\delta(\epsilon))}\label{eq:e4s9}
  \end{IEEEeqnarray}
  where \eqref{eq:e4s1} is due to the union bound, \eqref{eq:e4s2} and \eqref{eq:e4s3} respectively follow from the law of total probability and the chain rule, \eqref{eq:e4s4} is due to the Markov chain relationships in \eqref{eq:mc1} and \eqref{eq:mc2}, the inequality in \eqref{eq:e4s5} is obtained using\cite[Lemma 1]{kim2015} with the correspondences 
  \[S \leftrightarrow \tilde{S}^{(B)}_1, U \leftrightarrow \tilde{U}^{(B)}_1, \epsilon'\leftrightarrow\epsilon_1, \text{\ and\ } M \leftrightarrow M^{(B)}_1,\] \eqref{eq:e4s7}-\eqref{eq:e4s9} follow standard bounds for typical sets, and in the last equation we set $\delta(\epsilon)\triangleq\delta_1(\epsilon)+\delta_2(\epsilon)$.\footnote{Note that $\lim_{\epsilon\to 0}\delta_1(\epsilon)=0$ and $\lim_{\epsilon\to 0}\delta_2(\epsilon)=0$.} 
  Therefore, if 
  \[R_1^{(B)}<I(\tilde{U}_1; S_2, U_2, \tilde{S}_2, \tilde{U}_2, \tilde{W}_2, X_2, Y_2)-\delta(\epsilon)\] holds, then $\lim_{n\to\infty}\Pr(\mathcal{F}_4^{(B+1)})=0$. 
  By symmetry, one can easily obtain a similar condition for terminal~$2$. 
  Combining the first part then completes the proof. 
\end{IEEEproof}\medskip

\noindent \textbf{Claim~7}: For $b=2, 3, \dots, B$, if $R^{(b)}_j > I(S_j; U_j)+\delta_1(\epsilon_1)$ and $R^{(b-1)}_1<I(\tilde{U}_1; S_{2}, U_{2}, \allowbreak\tilde{S}_{2}, \allowbreak\tilde{U}_{2}, \allowbreak\tilde{W}_{2}, X_{2}, Y_{2})-\delta(\epsilon)$, then $\lim_{n\to\infty}\Pr\big(\mathcal{E}_{1}^{(b)}\cap\overline{\mathcal{E}}_{1}^{(b-1)}\big)=0$.
\begin{IEEEproof}
  We sketch the proof since the details follow similar lines of the proofs for Claims~5 and~6.
  Using Claim~3, it suffices to show that under the hypothesis, we have that $\lim_{n\to\infty}\Pr(\mathcal{F}^{(b)}_j)\allowbreak =0$ for $j=1, 2$, $\lim_{n\to\infty}\Pr(\overline{\mathcal{F}^{(b)}_1\cup\mathcal{F}^{(b)}_2}\cap\mathcal{F}_3^{(b)}\cap\overline{\mathcal{E}}_1^{(b-1)})=0$, and $\lim_{n\to\infty}\Pr(\mathcal{F}_4^{(b)})=0$. 
  Note that the first two quantities can be easily proved using the argument in the first part of the proof for Claim~5, which imposes the condition $R^{(b)}_j>I(S_j; U_j)+\delta(\epsilon_1)$ for $j=1, 2$. 

  To show $\lim_{n\to\infty}\Pr(\overline{\mathcal{F}^{(b)}_1\cup\mathcal{F}^{(b)}_2}\cap\mathcal{F}_3^{(b)}\cap\overline{\mathcal{E}}_1^{(b-1)})=0$, we follow the proofs of Claim~5 and~6. 
  Based on the proof of Claim~5, it is straightforward to obtain that $\lim_{n\to\infty}\Pr(\overline{\mathcal{F}^{(b)}_1\cup\mathcal{F}^{(b)}_2})=1$ under the conditions $R^{(b)}_j>I(S_j; U_j)+\delta(\epsilon_1)$, $j=1, 2$. 
  Consider the inequality $\Pr(\overline{\mathcal{F}^{(b)}_1\cup\mathcal{F}^{(b)}_2}\cap\mathcal{F}_3^{(b)}\cap\overline{\mathcal{E}}_1^{(b-1)})\le\Pr(\mathcal{F}_3^{(b)}|\overline{\mathcal{F}^{(b)}_1\cup\mathcal{F}^{(b)}_2}\cap\overline{\mathcal{E}}_1^{(b-1)})$, where the event $\overline{\mathcal{E}}_1^{(b-1)}$ implies that $(\tilde{\bm{S}}_1^{(b)}, \tilde{\bm{S}}_2^{(b)}, \tilde{\bm{U}}_1^{(b)}, \allowbreak\tilde{\bm{U}}_2^{(b)}, \allowbreak\tilde{\bm{W}}_1^{(b)}, \allowbreak\tilde{\bm{W}}_2^{(b)})$ is a jointly typical sequence. 
  Noting that the right-hand-side of the inequality is now at a position similar to $\Pr(\mathcal{F}_3^{(B+1)}|\overline{\mathcal{E}}_1^{(B)})$ in the proof of Claim~6, we obtain the desired result by applying conditional typicality lemma twice as done before. 

  For the probability $\Pr(\mathcal{F}_4^{(b)})$, we adopt the proof of Claim~6 with the correspondence $B+1\leftrightarrow b$, which imposes the sufficient condition $R^{(b-1)}_1<I(\tilde{U}_1; \allowbreak S_{2}, \allowbreak U_{2}, \allowbreak\tilde{S}_{2}, \allowbreak\tilde{U}_{2}, \tilde{W}_{2}, X_{2}, Y_{2})-\delta(\epsilon)$ for $\lim_{n\to\infty}\Pr(\mathcal{F}_4^{(b)})=0$. 
  Combining the above results then completes the proof. 
\end{IEEEproof}

\subsection{Auxiliary Result for Special Case (ii) of Corollary~\ref{cor:twchybrid}}\label{subsec:scii}
By symmetry, we only show that $I(\tilde{S}_1; \tilde{U}_1|\tilde{S}_2, \tilde{U}_2) < I(\tilde{U}_1; Y_2|\tilde{S}_2, \tilde{U}_2)$ reduces to $R^{(1)}(D_1)<I(X_1; Y_2|X_2)$. 
First, observe that
\begin{IEEEeqnarray}{rCl}
  I(\tilde{S}_1; \tilde{U}_1|\tilde{S}_2, \tilde{U}_2)&=&I(\tilde{S}_1; V'_1, \hat{S}'_1|\tilde{S}_2, V'_2, \hat{S}'_2)\nonumber\\
  &=& \underbrace{I(\tilde{S}_1; V'_1|\tilde{S_2}, V'_2, \hat{S}'_2)}_{=0}+I(\tilde{S}_1; \hat{S}'_1|\tilde{S}_2, V'_2, \hat{S}'_2, V'_1)\nonumber\\
  &=& H(\hat{S}'_1|\tilde{S}_2, V'_2, \hat{S}'_2, V'_1)-H(\hat{S}'_1|\tilde{S}_2, V'_2, \hat{S}'_2, V'_1, \tilde{S}_1)\nonumber\\
  &=& H(\hat{S}'_1)-H(\hat{S}'_1|\tilde{S}_1)\label{eq:sp1}\\
  &=& I(\tilde{S}_1; \hat{S}'_1)\nonumber\\
  &=& R^{(1)}(D_1)\label{eq:sp2}
\end{IEEEeqnarray}
where \eqref{eq:sp1} holds since $\tilde{S}_1$ and $\tilde{S}_2$ are independent and hence $\hat{S}'_1$ is independent of $(\tilde{S}_2, V'_2, \hat{S}'_2, V'_1)$, and \eqref{eq:sp2} follows since the joint probability distribution $P_{\tilde{S}_1, \hat{S}'_1}=P_{S_1, \hat{S}_1}$ achieves $R^{(1)}(D_1)$. 

Moreover, we have that
\begin{IEEEeqnarray}{rCl}
  I(\tilde{U}_1; Y_2|\tilde{S}_2, \tilde{U}_2)&=&I(V'_1, \hat{S}'_1; Y_2|\tilde{S}_2, V'_2, \hat{S}'_2)\nonumber\\
  &=& I(V'_1; Y_2|\tilde{S}_2, V'_2, \hat{S}'_2)+I(\hat{S}'_1; Y_2|\tilde{S}_2, V'_2, \hat{S}'_2, V'_1)\nonumber\\
  &=& I(X_1; Y_2|\tilde{S}_2, X_2, \hat{S}'_2)+I(\hat{S}'_1; Y_2|\tilde{S}_2, X_2, \hat{S}'_2, X_1)\label{eq:sp3}\\
  &=& H(Y_2|\tilde{S}_2, X_2, \hat{S}'_2)-H(Y_2|\tilde{S}_2, X_2, \hat{S}'_2, X_1)\label{eq:sp4}\\
  &=& H(Y_2|X_2)-H(Y_2|X_2, X_1)\label{eq:sp5}\\
  &=& I(X_1; Y_2|X_2)\nonumber
\end{IEEEeqnarray}
where \eqref{eq:sp3} follows since $X_j=V'_j$, \eqref{eq:sp4} holds since given channel inputs $X_1$ and $X_2$, the output $Y_2$ is independent of other variables, and \eqref{eq:sp5} holds due to the Markov chain relationship $(\tilde{S}_2, \hat{S}'_2)\markov X_2\markov Y_2$. 

\subsection{Proof of Converse Part in Theorem~\ref{thm:JSCC4}}\label{sub:jscc4converse}
For $k_1\le k_2$, let $S_{j, k_1}^{k_2}\triangleq (S_{j, k_1}, S_{j, k_1+1}, \dots, S_{j, k_2})$. 
Given a rate-$K/N$ joint source-channel code that achieves the distortion pair $(D_1, D_2)$, we obtain \eqref{eq:OB3a} by the following derivation: 
{\allowdisplaybreaks
\begin{IEEEeqnarray}{rCl}
K\cdot R_{S_1|S_0}(D_1)&\le &K\cdot R_{S_1|S_0}\left(K^{-1}\sum\limits_{k=1}^K \mathbb{E}\left[d_1(S_{1, k}, \hat{S}_{1, k})\right]\right)\label{eq:cc-0}\\
&\le &\sum\limits_{k=1}^K R_{S_1|S_0}\left(\mathbb{E}[d_1(S_{1, k}, \hat{S}_{1, k})]\right)\label{eq:cc-1}\\
&\le &\sum\limits_{k=1}^K I(S_{1, k}; \hat{S}_{1, k}|S_{0, k})\label{eq:cc-2}\\
&\le &\sum\limits_{k=1}^K I(S_{1, k}; S_2^K, Y_2^N|S_{0, k})\label{eq:cc-3}\\
&\le & \sum\limits_{k=1}^K H(S_{1, k}|S_{0, k})-H(S_{1, k}|S_{0}^k, S_2^K, Y_2^N)\label{eq:cc-8}\\
&\le &\sum\limits_{k=1}^K H(S_{1, k}|S_0^K, S_1^{k-1}, S_2^K)-H(S_{1, k}|S_0^K, S_1^{k-1}, S_{2}^K, Y_2^N)\label{eq:cc-4}\IEEEeqnarraynumspace\\
&= & \sum\limits_{k=1}^K I(S_{1, k}; Y_2^N|S_0^K, S_1^{k-1}, S_2^K)\nonumber\label{eq:cc-9}\\
&= & I(S_1^K; Y_2^N|S_0^K, S_2^K)\nonumber\\
&= & \sum_{n=1}^N I(S_1^K; Y_{2, n}|S_0^K, S_2^K, Y_2^{n-1})\nonumber\\
&\le &\sum_{n=1}^N H(Y_{2, n}|X_{2, n})-H(Y_{2, n}|S_0^K, S_1^K, S_2^K, Y_2^{n-1}, X_{1, n}, X_{2, n})\label{eq:cc-5}\\
&= &\sum_{n=1}^N H(Y_{2, n}|X_{2, n})-H(Y_{2, n}|X_{1, n}, X_{2, n})\label{eq:cc-7}\\
&= & N\cdot\sum\limits_{n=1}^N \frac{1}{N}\cdot I(X_{1, n}; Y_{2, n}|X_{2, n})\nonumber\\
&\le & N\cdot I(X_1; Y_2|X_2)\label{eq:cc-6},
\end{IEEEeqnarray}
where \eqref{eq:cc-0} holds since $R_{S_1|S_0}(D_1)$ is non-increasing and the expected distortion of the code is not larger than $D_1$, \eqref{eq:cc-1} and \eqref{eq:cc-2} are respectively due to convexity and the definition of conditional RD function, \eqref{eq:cc-3} follows from the data-processing inequality, \eqref{eq:cc-8} holds since conditioning reduces entropy, \eqref{eq:cc-4} holds by the Markov chain relationships $S_{1, k}\markov S_{0, k}\markov (S_0^{k-1}, S_{0,k+1}^K, S_1^{k-1})$ and $S_1^K\markov S_0^K\markov S_2^K$ and since conditioning reduces entropy, \eqref{eq:cc-5} holds since $X_{2, n}$ is a function of $(Y_2^{n-1}, S_2^K)$ and since conditioning reduces entropy, \eqref{eq:cc-7} follows from the memoryless property of channel, and \eqref{eq:cc-6} holds with $P_{X_1, X_2}=N^{-1}\sum_{n=1}^N P_{X_{1, n}, X_{2, n}}$ since $I(X_{1, n}; Y_{2, n}|X_{2, n})$ is concave in $P_{X_{1, n}, X_{2, n}}$.  
By symmetry, a similar argument shows \eqref{eq:OB3b}. \hfill \IEEEQEDhere
}

\end{appendix}


\end{document}